\documentclass[a4paper,fleqn,usenatbib]{mnras}
\pdfoutput=1
\usepackage{mathptmx}

\usepackage[T1]{fontenc}
\usepackage{ae,aecompl}

\usepackage{graphicx}	
\usepackage{amsmath}	
\usepackage{amssymb}	
\usepackage{listings}
\usepackage{color}

\definecolor{codegreen}{rgb}{0,0.6,0}
\definecolor{codegray}{rgb}{0.5,0.5,0.5}
\definecolor{codepurple}{rgb}{0.58,0,0.82}
\definecolor{backcolour}{rgb}{0.95,0.95,0.92}

\newcommand{\dnu}{$\Delta\nu$}
\newcommand{\numax}{$\nu_{\rm max}$}

\lstdefinestyle{mystyle}{
  backgroundcolor=\color{backcolour},   commentstyle=\color{codegreen},
  keywordstyle=\color{magenta},
  numberstyle=\tiny\color{codegray},
  stringstyle=\color{codepurple},
  basicstyle=\footnotesize,
  breakatwhitespace=false,         
  breaklines=true,                 
  captionpos=b,                    
  keepspaces=true,                 
  numbers=left,                    
  numbersep=5pt,                  
  showspaces=false,                
  showstringspaces=false,
  showtabs=false,                  
  tabsize=2
}
\title[The mass dependence in the $\Delta\nu$ scaling relation]{Mitigating the mass dependence in the $\Delta\nu$ scaling relation of red-giant stars}

\author[E. Guggenberger et al.]{
Elisabeth Guggenberger,$^{1,2}$\thanks{E-mail: guggenberger@mps.mpg.de}
Saskia Hekker,$^{1, 2}$
George C.~Angelou,$^{1, 2}$
\newauthor \hspace{0.25mm} Sarbani Basu,$^{3}$
 Earl P.~Bellinger$^{1, 2, 3, 4}$ 
\\
$^{1}$Max Planck Institut f\"ur Sonnensystemforschung, Justus-von-Liebig-Weg 3, 37077 G\"ottingen, Germany\\
$^{2}$Stellar Astrophysics Centre, Dept. of Physics and Astronomy, Aarhus University, Ny Munkegade 120, 8000 Aarhus C, Denmark\\
$^{3}$Department of Astronomy, Yale University, 52 Hillhouse Avenue, New Haven, CT 06511, USA\\
$^{4}$Institut f\"ur Informatik, Georg-August-Universit\"at G\"ottingen, Goldschmidtstrasse 7, D-37077 G\"ottingen, Germany
}

\date{Accepted XXX. Received YYY; in original form ZZZ}

\pubyear{2015}

\begin{document}
\label{firstpage}
\pagerange{\pageref{firstpage}--\pageref{lastpage}}
\maketitle

\begin{abstract}
The masses and radii of solar-like oscillators can be estimated through the asteroseismic scaling relations. These relations provide a direct link between observables, i.e. effective temperature and characteristics of the oscillation spectra, and stellar properties, i.e. mean density and surface gravity (thus mass and radius). These scaling relations are commonly used to characterize large samples of stars.
Usually, the Sun is used as a reference from which the structure is scaled. However, for stars that do not have a similar structure as the Sun, using the Sun as a reference introduces systematic errors as large as 10\% in mass and 5\% in radius. Several alternatives for the reference of the scaling relation involving the large frequency separation (typical frequency difference between modes of the same degree and consecutive radial order) have been suggested in the literature. In a previous paper, we presented a reference function with a dependence on both effective temperature and metallicity. The accuracy of predicted masses and radii improved considerably when using reference values calculated from our  reference function. However, the residuals indicated that stars on the red-giant branch possess a mass dependence that was not accounted for. Here, we present a reference function for the scaling relation involving the large frequency separation that includes the mass dependence. This new reference function improves the derived masses and radii significantly by removing the systematic differences and mitigates the trend with $\nu_{\rm max}$ (frequency of maximum oscillation power) that exists when using the solar value as a reference.

\end{abstract}

\begin{keywords}
stars: fundamental parameters -- asteroseismology -- stars: general -- stars: oscillations
\end{keywords}



\section{Introduction}

\begin{figure*}
\centering
\begin{minipage}{0.45\linewidth}
 \includegraphics[width=\linewidth]{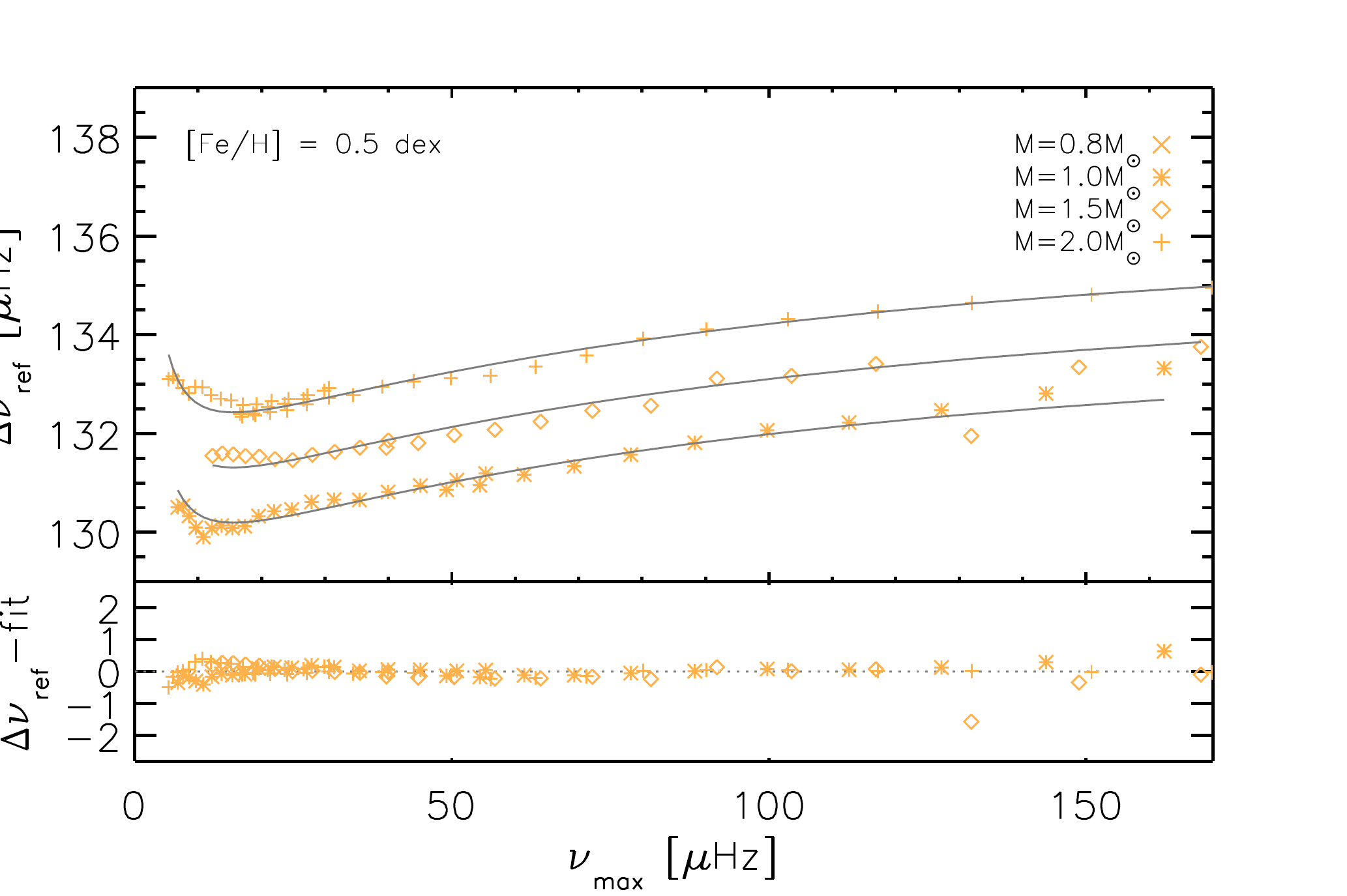}
\end{minipage}
\begin{minipage}{0.45\linewidth}
 \includegraphics[width=\linewidth]{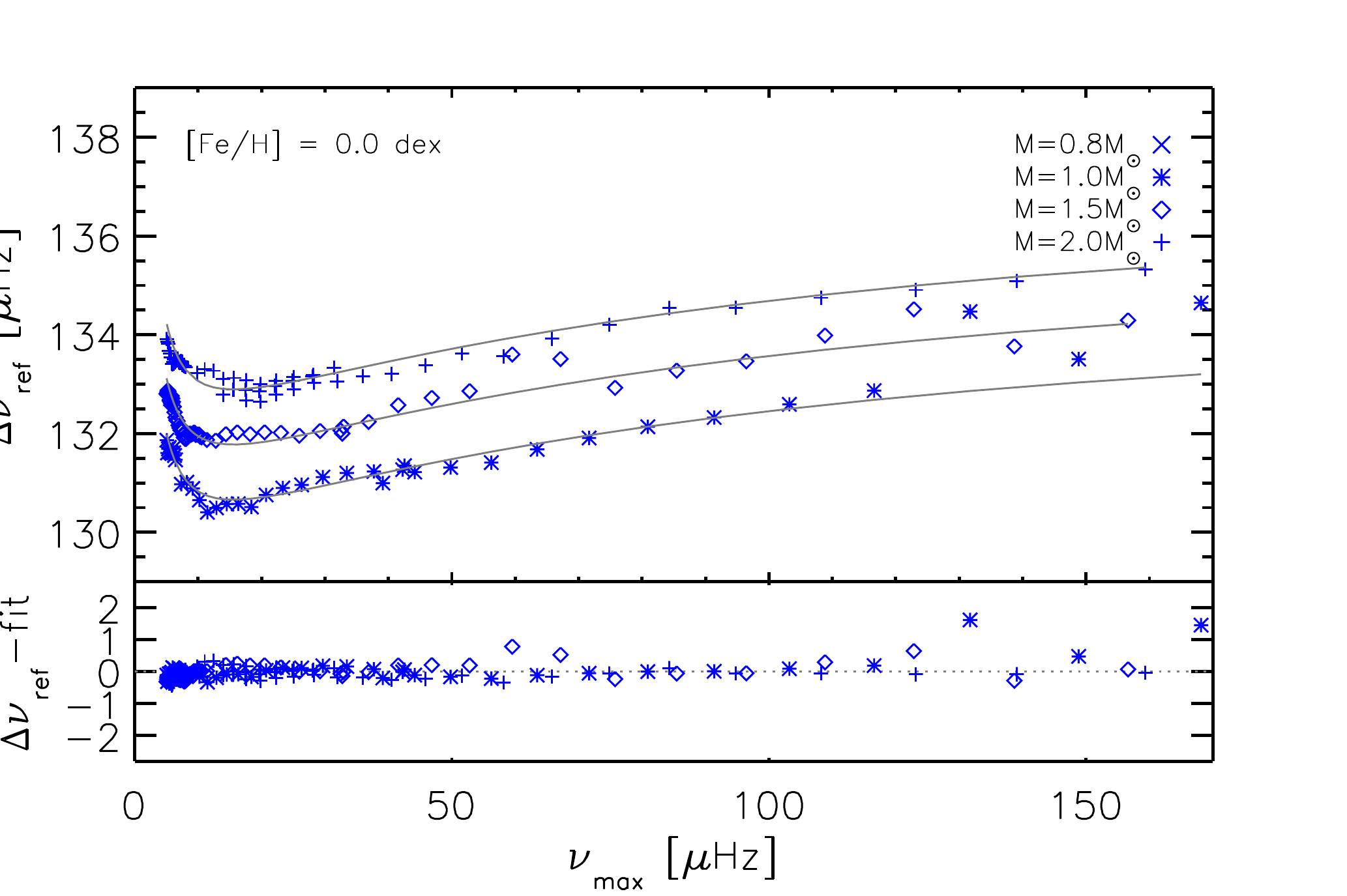}
\end{minipage}
\begin{minipage}{0.45\linewidth}
 \includegraphics[width=\linewidth]{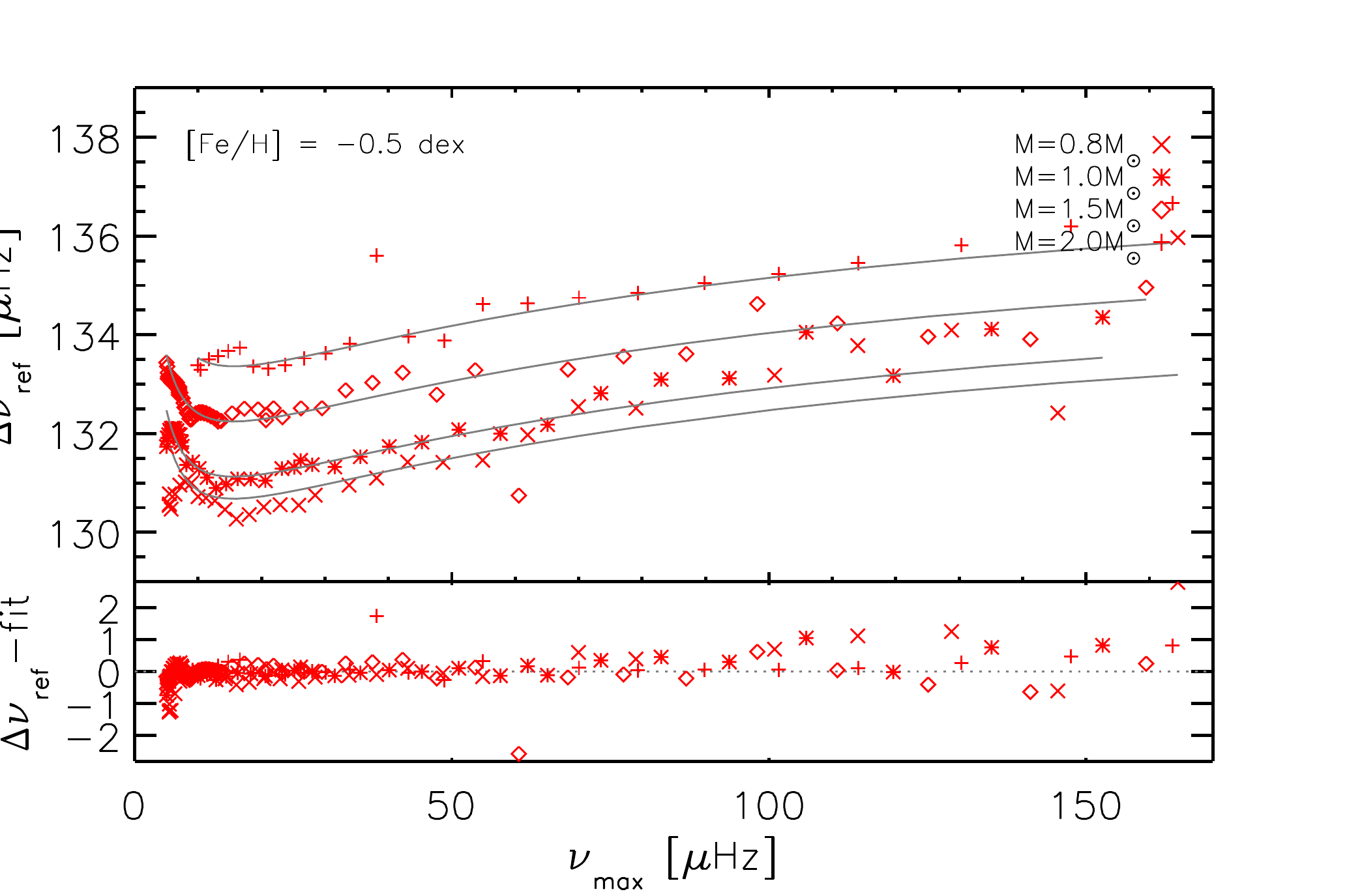}
\end{minipage}
\begin{minipage}{0.45\linewidth}
 \includegraphics[width=\linewidth]{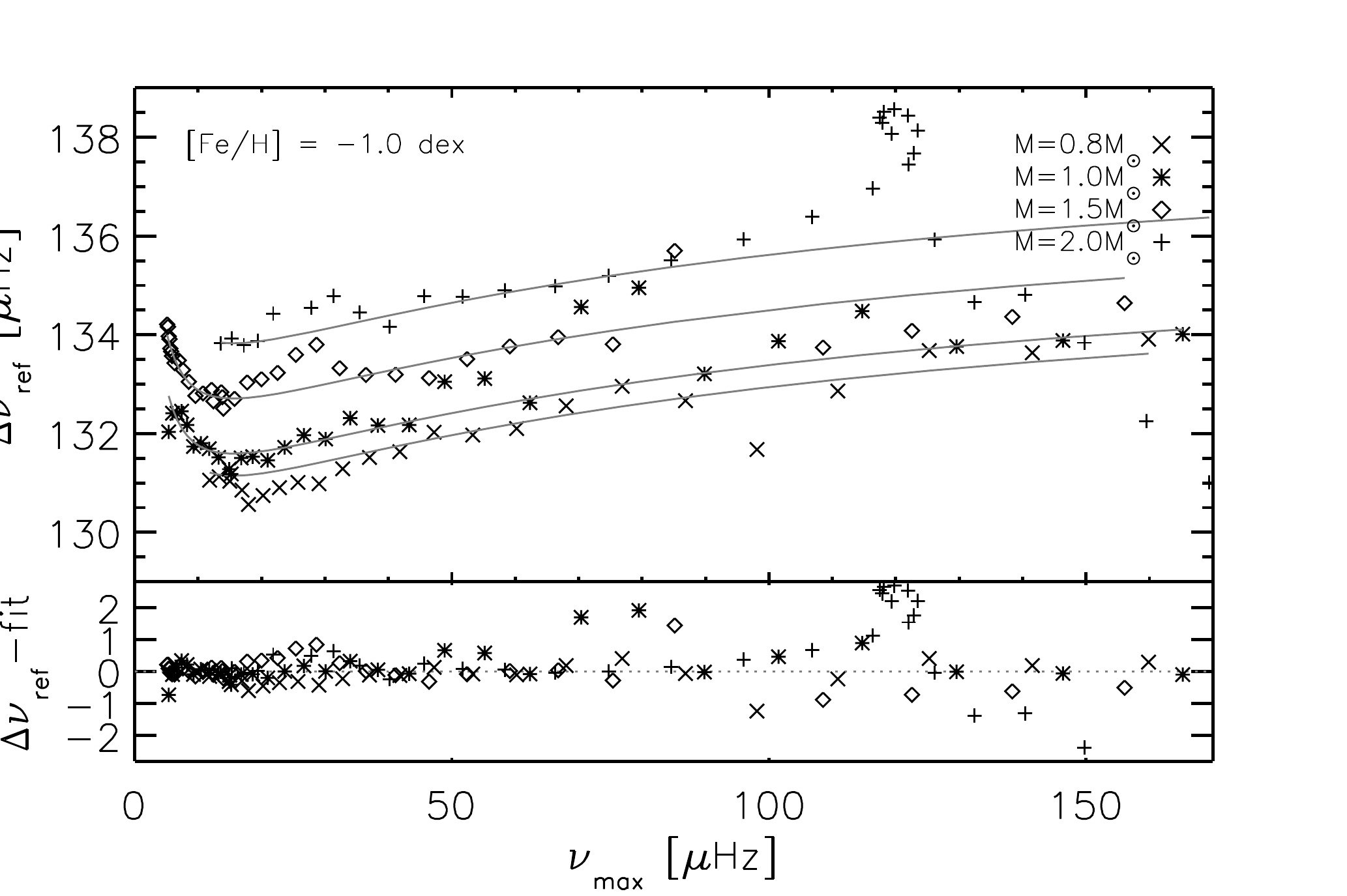}
\end{minipage}
 \caption{Reference values for \dnu\ ($\Delta\nu_{\rm ref}$) as a function of \numax\ with in grey the mass dependent fit (Eq.~\ref{eq1}). Each panel shows one metallicity and different masses with symbols indicated in the legend. Residuals are shown in the lower part of each panel with the grey dotted line indicating zero.}
 \label{fig:fit}
\end{figure*}

The asteroseismic scaling relations \citep{ulrich86, kb95} permit a straightforward inference of 
 fundamental stellar parameters such as the mass and radius of solar-like oscillators (i.e., stars that exhibit stochastic oscillations driven by turbulent convection in their outer layers). These relations require knowledge of the effective temperature and some characteristics of the oscillations that can be derived from timeseries data of either brightness or radial velocity variations. One of these oscillation characteristics is the so-called large frequency separation, \dnu, which is defined as the separation between modes of the same degree and consecutive radial orders. The other oscillation parameter is the frequency of maximum oscillation power, \numax. These scaling relations relate \dnu\ and \numax\ to mean density $\overline{\rho}$ and surface gravity $g$, and thus mass $M$ and radius $R$, in the following way:


\begin{equation}
{\Delta\nu} \simeq \sqrt{\overline{\rho}}\, \Delta \nu_{\rm ref} = \sqrt{\frac{M}{R^3}}\, \Delta\nu_{\rm ref},
\label{dnu}
\end{equation}
\begin{equation} 
\nu_{\rm{max}}\simeq \frac{g}{\sqrt{T_{\rm eff}}} \, \nu_{\rm max,ref} =  \frac{M}{R^2\sqrt{T_{\rm{eff}}}}\, \nu_{\rm max,ref}.
\label{numax}
\end{equation}
 where $\overline{\rho}$, $g$, $M$, $R$ and effective temperature $T_{\rm eff}$ are expressed in solar values. In the initial formulations of these relations the solar quantities are used as reference values, i.e. \dnu$_{\rm ref}$~=~\dnu$_\odot$~=~135.1~$\pm$~0.1~$\mu$Hz, \citep{hub11}, and $\nu_{\rm max,ref}$~=~\numax$_\odot$~=~3050~$\mu$Hz \citep{kb95}.
 This approach inherently assumes that the Sun and the star under investigation have similar internal structures. However this assumption is not strictly valid and differences in stellar structure cause systematic errors in the masses and radii \citep[see e.g.][]{corsaro13,hub13, ep14, mig12} derived from the scaling relations.
Several approaches to correct for the errors have been suggested, including interpolations between precomputed models \citep{sharma16}, which requires access to a model grid, as well as analytical formulations \citep{white11,moss13}. For a detailed discussion of the available corrections, see \citet[][hereinafter Paper~I]{gug16}.  

In Paper~I we used stellar models to derive a reference large frequency separatio $\Delta\nu_{\rm ref}$ that should be used instead of $\Delta\nu_{\odot}$. We modelled the general behaviour of $\Delta\nu_{\rm ref}$ over a wide range of temperatures and metallicities with a damped sinusoid, thereby extending the work of \citet{white11} to a larger parameter space. The analytical expression presented in Paper~I can be applied in a straight-forward manner to compute the optimal reference value $\Delta\nu_{\rm ref}$ from the observables T$_{\rm eff}$ and [Fe/H] and reduces the errors in masses and radii by a factor of two compared to the uncorrected scaling relations. However, on the red-giant branch a dependence on mass remained present in the residuals of $\Delta\nu_{\rm ref}$ (see Fig. 3 of Paper~I). The dispersion of the different mass tracks occurs at different temperatures depending on metallicity, ranging from about 4500\,K at [Fe/H] = 0.5 to about 5200\,K at [Fe/H] = -1.0. In all cases these values correspond to the same evolutionary phase; namely, where the star turns onto the red-giant branch (see e.g. Fig. 2 of Paper~I).
In the current work we mitigate this mass dependence on the red-giant branch. 

\section{Methods}
\label{sec:methods} 

\subsection{Models and \texorpdfstring{$\Delta\nu_{\rm ref}$}{reference large frequency separation} determination }
Stellar models were calculated with the the Yale Rotating Stellar Evolution Code \citep[YREC,][]{dem08}. The choice of input physics are as described in paper~I with our grid comprising masses of 0.8, 1.0, 1.5 and 2.0 M$_\odot$ and metallicities [Fe/H] = 0.5, 0.0, -0.5 and -1.0 dex. A subset of models along each track were extracted for the analysis such that  $\nu_{\rm max} > 5$~$\mu$Hz. 

The value of \dnu\ was determined from the models using radial mode frequencies as per Paper~I, i.e. using a Gaussian weighted linear fit to the frequencies around the expected \numax, where the full-width at half maximum of the Gaussian is fixed to 5\,\dnu. From this value of \dnu, together with the known model mass and radius, we computed the reference $\Delta\nu_{\rm{ref}}$ needed to reproduce the model mass and radius from the \dnu\ scaling relation (see Figs~\ref{fig:fit} and \ref{fig:fit_a}). We note that the scatter in $\Delta\nu_{\rm ref}$ increases as a function of decreasing [Fe/H]. The reason for this is currently unknown but this may be due to difficulties in the computation of the mode frequencies. Uncertainties in the frequencies also set the lower limit of the $\nu_{\rm max}$ range that we consider.

Finally, we determined an analytical expression that allows the reference value to be calculated straightforwardly for use by the community.



\begin{figure*}
\centering
\begin{minipage}{0.45\linewidth}
 \includegraphics[width=\linewidth]{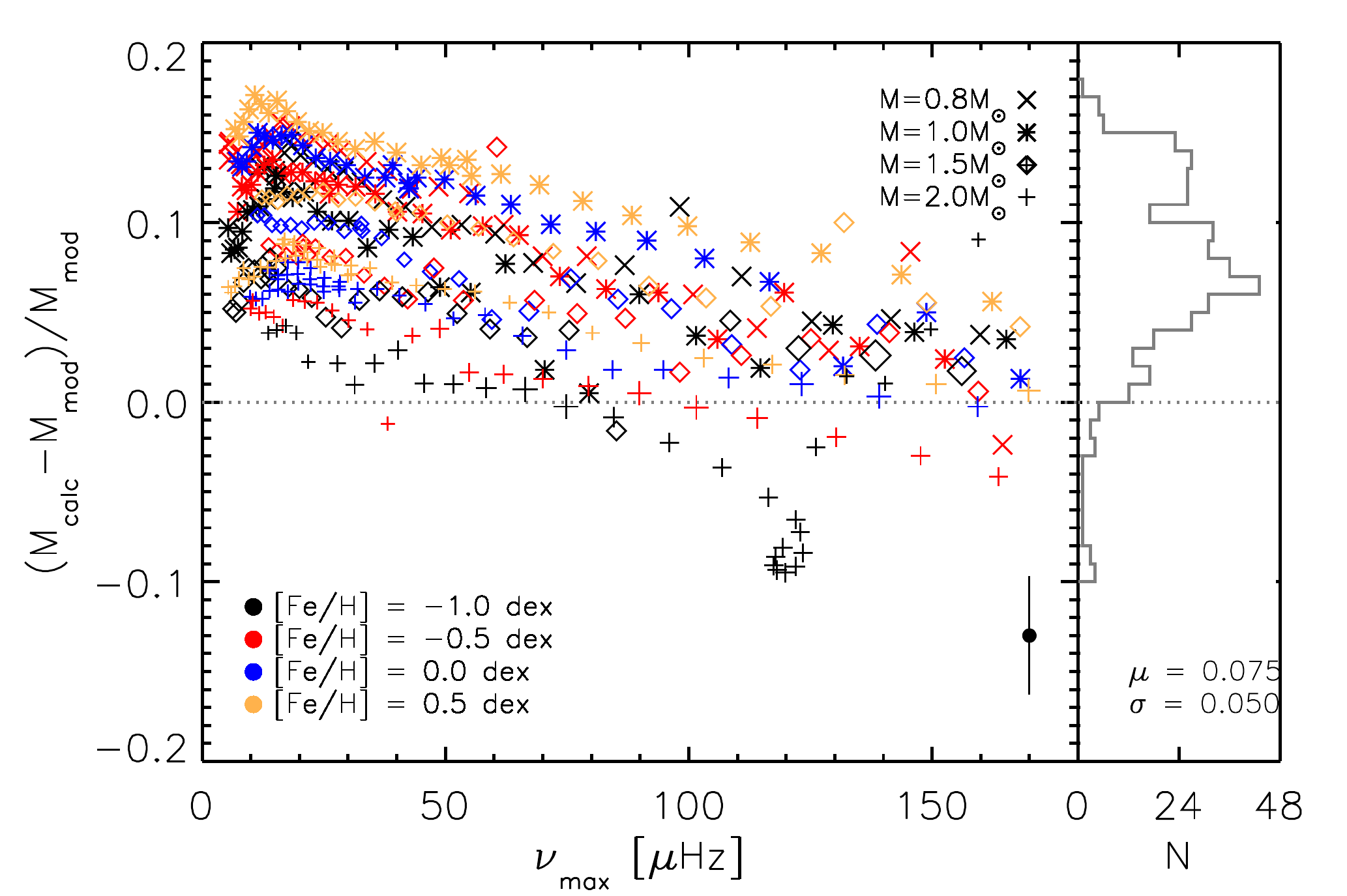}
\end{minipage}
 \begin{minipage}{0.45\linewidth}
 \includegraphics[width=\linewidth]{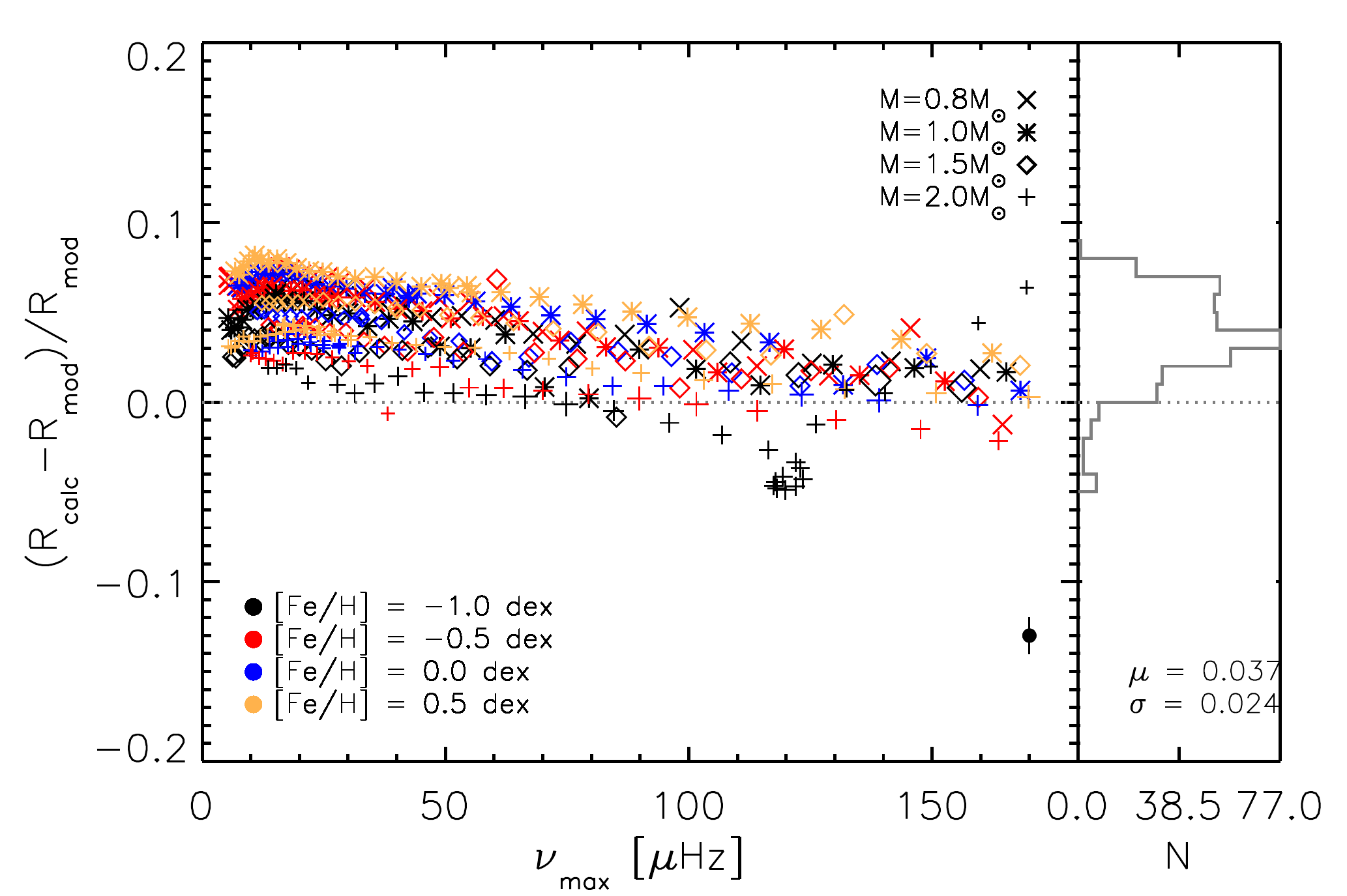}
\end{minipage}
\begin{minipage}{0.45\linewidth}
 \includegraphics[width=\linewidth]{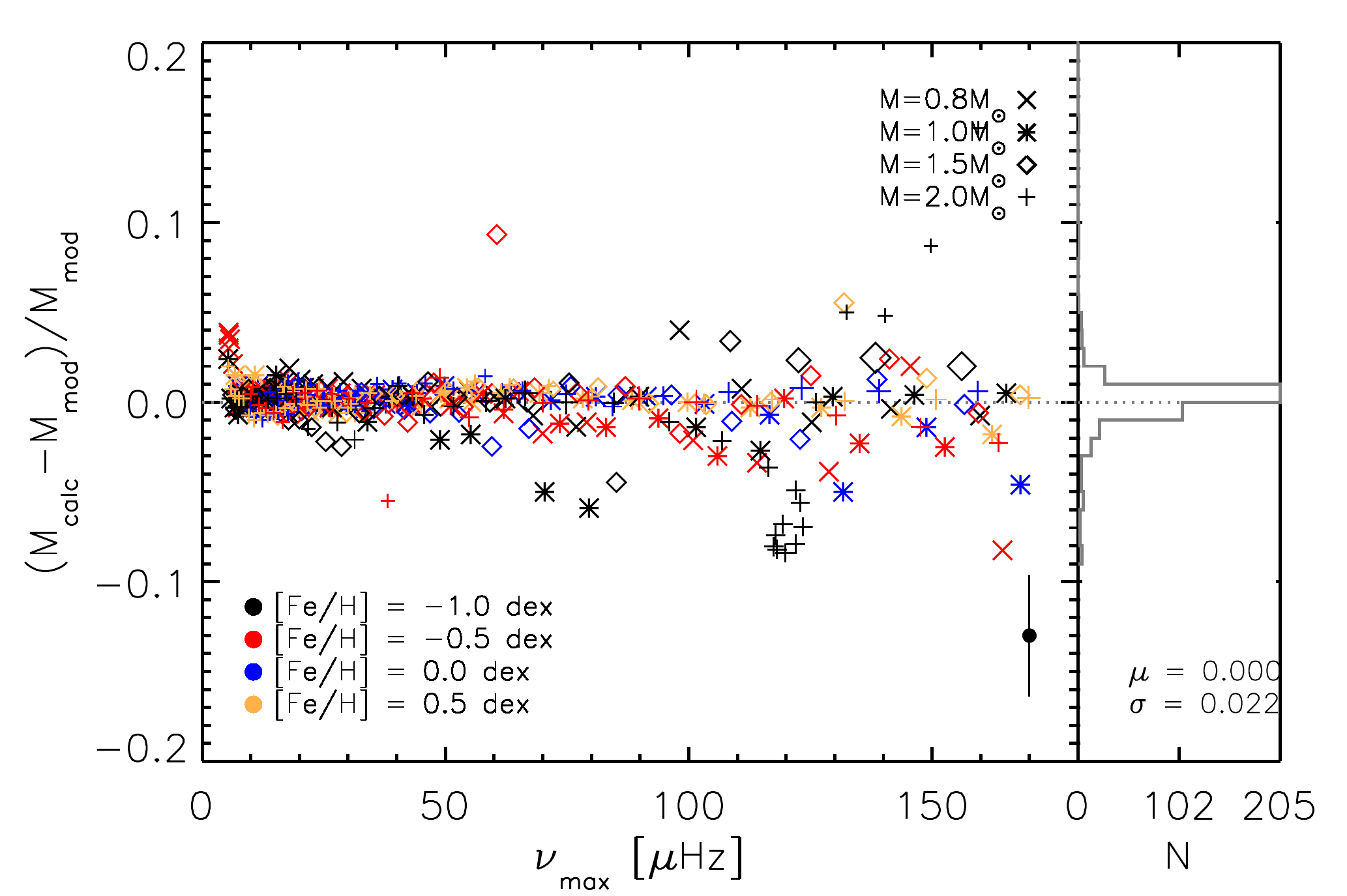}
\end{minipage}
 \begin{minipage}{0.45\linewidth}
 \includegraphics[width=\linewidth]{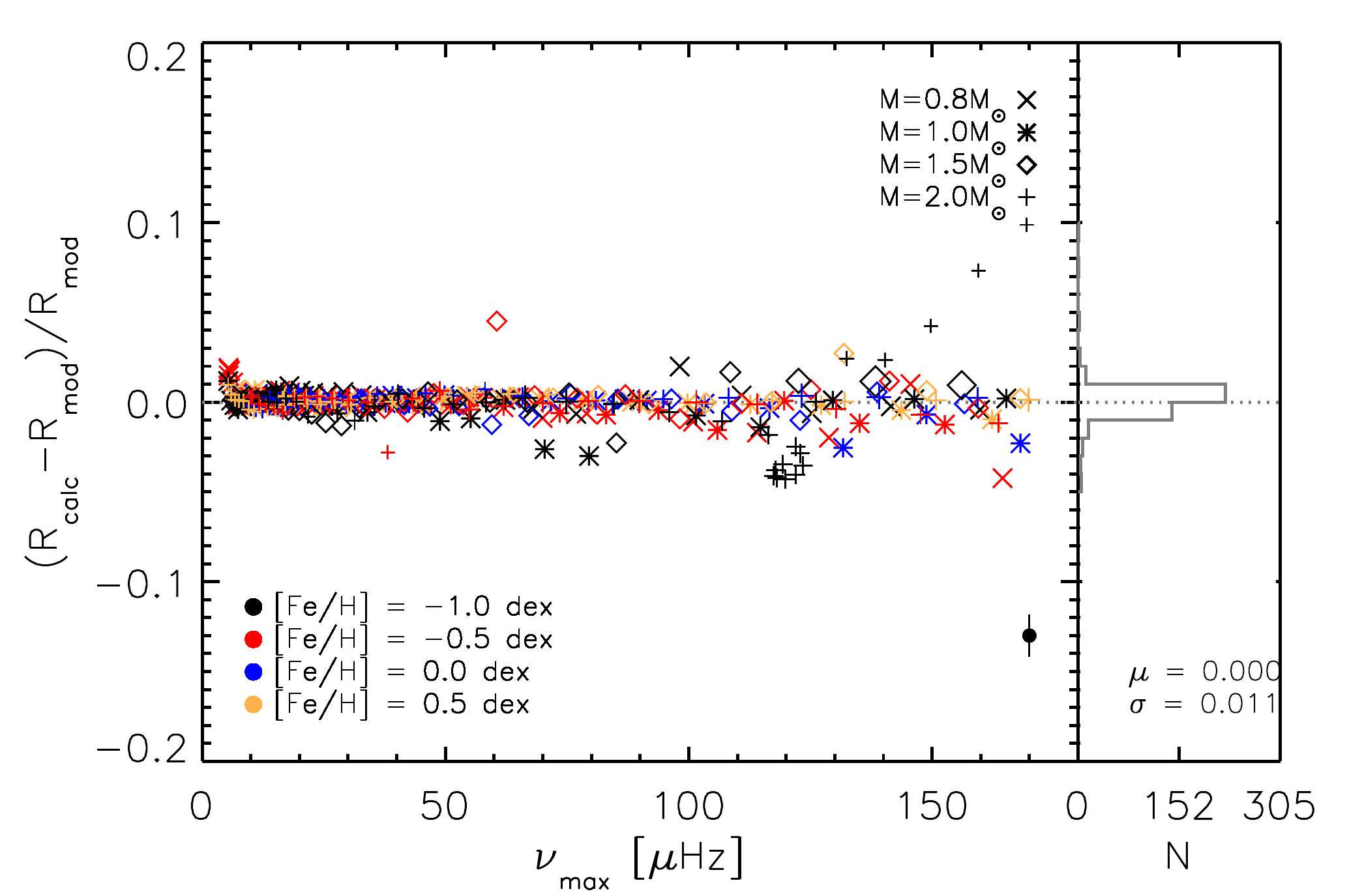}
\end{minipage}
 \caption{Mass (left) and radius (right) computed using $\Delta\nu_{\odot}$ (top) and Eq.~\ref{eq1} (bottom) as a reference for the \dnu\ scaling relation. The meaning of the colours and symbols are indicated in the legends. The dot with error bars in the right bottom corner of each panel shows a typical uncertainty computed as the standard deviation of the values obtained in the Monte Carlo test divided by the known value of the parameter. The symbol size is indicative of the number of iterations used with larger symbols indicating less iterations. The number of iterations ranges from 2-5. The histograms on the right side of each panel show the distribution of the results. The mean ($\mu$) and standard deviation ($\sigma$) of the relative differences are indicated in the legend. The gray dotted line marks zero.}
 \label{fig:MR}
\end{figure*}

\subsection{Symbolic regression}
\label{sec:symreg} 
In Paper~I we found that the underlying functional form of the reference \dnu\ as a function of $T_{\rm eff}$ resembles a damped sinusoid. The metallicity dependence was then included by investigating how the individual constants of the damped sinusoid behaved when the other  parameters of the function were kept fixed. The mass dependence investigated here shows a more complex behaviour. Therefore, we used symbolic regression. Symbolic regression is a type of optimisation that searches mathematical expressions to find a model that best fits the data. Thus rather than searching for the optimal parameters of a pre-defined expression (as is commonly done in regression techniques), the algorithm simultaneously optimizes the functional form  and the parameters of the function.  

For the work presented here we used the software \textsc{Eureqa} \citep{schmidt09} distributed by Nutonian. This software package uses an evolutionary search to automatically perform symbolic regression.



\section{Functions describing the mass dependence}
\label{sec:massdep} 

We performed symbolic regression on the $\Delta\nu_{\rm ref}$ values as derived in Paper~I as well as on the residuals that remain after applying the fit published in Paper~I ($\Delta\nu_{\rm ref,residuals}$). 
Our investigation revealed that a mass dependency in the correction function is possible if our expression takes the form $\Delta\nu_{\rm ref}$=$\Delta\nu_{\rm ref}$(\numax, M, [Fe/H]). 
We fit the function to stars  with   5~$\mu$Hz~$ <  \nu_{\rm max} < $~170~$\mu$Hz as the impact of the mass dependence is significant only in stars with \numax\ below 170~$\mu$Hz. For stars with higher \numax\ we therefore recommend the correction function from Paper~I.


The respective functions for \dnu$_{\rm{ref}}$ as well as for the residuals after the correction from Paper~I $\Delta\nu_{\rm ref,residuals}$ take the form:

\begin{multline}
\label{eq1}
\Delta\nu_{\mathrm{ref}} = \mathrm{A_1} + \mathrm{A_2} \cdot M + \frac{\mathrm{A_3}}{\nu_{\mathrm{max}}} + \\ \mathrm{A_4} \cdot \sqrt{\nu_{\mathrm{max}}} - \mathrm{A_5} \cdot {\nu_{\mathrm{max}}} - \mathrm{A_6} \cdot \mathrm{[Fe/H]}
\end{multline}

\begin{table}
	\centering
	\caption{Parameters with their units of the functions in Eqs~\ref{eq1} and \ref{eq2}.}
	\label{tab:param}
	\begin{tabular}{lrrclrr} 
		\hline
$A_1$	&	124.72	&  $\mu$Hz &	\  \	&	$B_1$	&	1.88	&  $\mu$Hz/M$_\odot$	\\
$A_2$	&	2.23	&  $\mu$Hz/M$_\odot$ &	\  \	&	$B_2$	&	0.02  &  -	\\	
$A_3$	&	17.61	&  $\mu$Hz$^2$ & \  \ &	$B_3$	&	5.14 & $\mu$Hz$^2/$M$_\odot$		\\
$A_4$	&	0.73	&  $\sqrt{\mu \rm Hz}$  & \  \ &	$B_4$	&	10.90 &  $\mu$Hz$^2$	\\
$A_5$	&	0.02	&  - &	\  \	&	$B_5$	&	3.69 &  $\mu$Hz		\\
$A_6$	&	0.93	&  $\mu$Hz &	\  \	&	$B_6$	&	0.01 & M$_{\odot}^{-1}$	\\	
		\hline
	\end{tabular}
\end{table}

\begin{multline}
\label{eq2}
\Delta\nu_{\mathrm{ref,residuals}} = \mathrm{B_1} \cdot M + \mathrm{B_2} \cdot \nu_{\mathrm{max}} + \\
\frac{\mathrm{B_3} \cdot M - \mathrm{B_4} \cdot \mathrm{[Fe/H]}}{\nu_{\mathrm{max}}} - \mathrm{B_5} - \mathrm{B_6} \cdot M \cdot \nu_{\mathrm{max}} 
\end{multline}
with the parameters and units listed in Table ~\ref{tab:param}.
We note that these reference functions (Eqs~\ref{eq1} and \ref{eq2}) as well as the reference function in Paper I do not have a clear physical meaning. They do represent however an empirical fit optimised to the data obtained from stellar models that include canonical stellar physics.

Example code for implementing the algorithm is provided in Appendix B.

\begin{figure*}
\centering
\begin{minipage}{0.45\linewidth}
 \includegraphics[width=\linewidth]{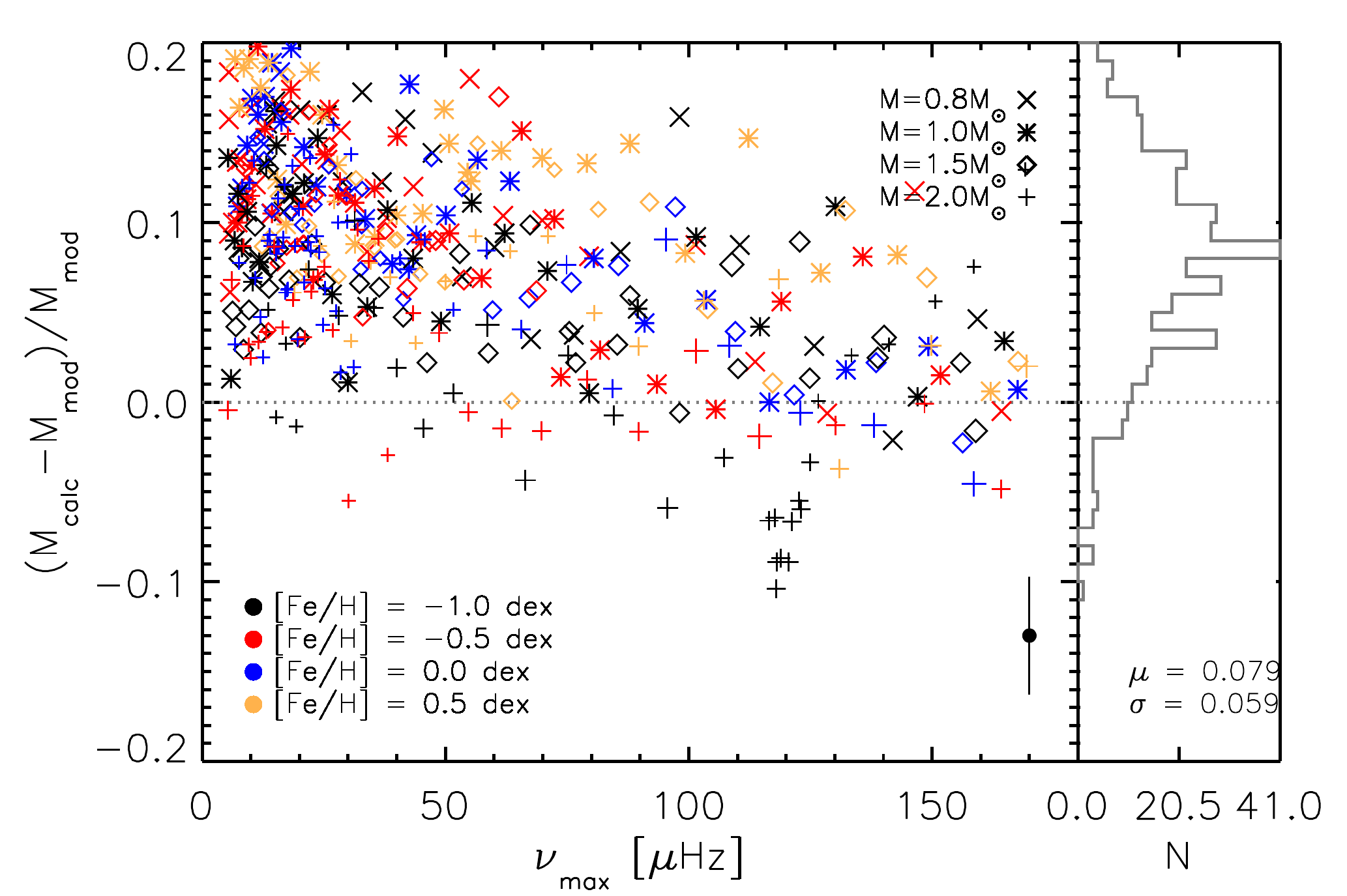}
\end{minipage}
 \begin{minipage}{0.45\linewidth}
 \includegraphics[width=\linewidth]{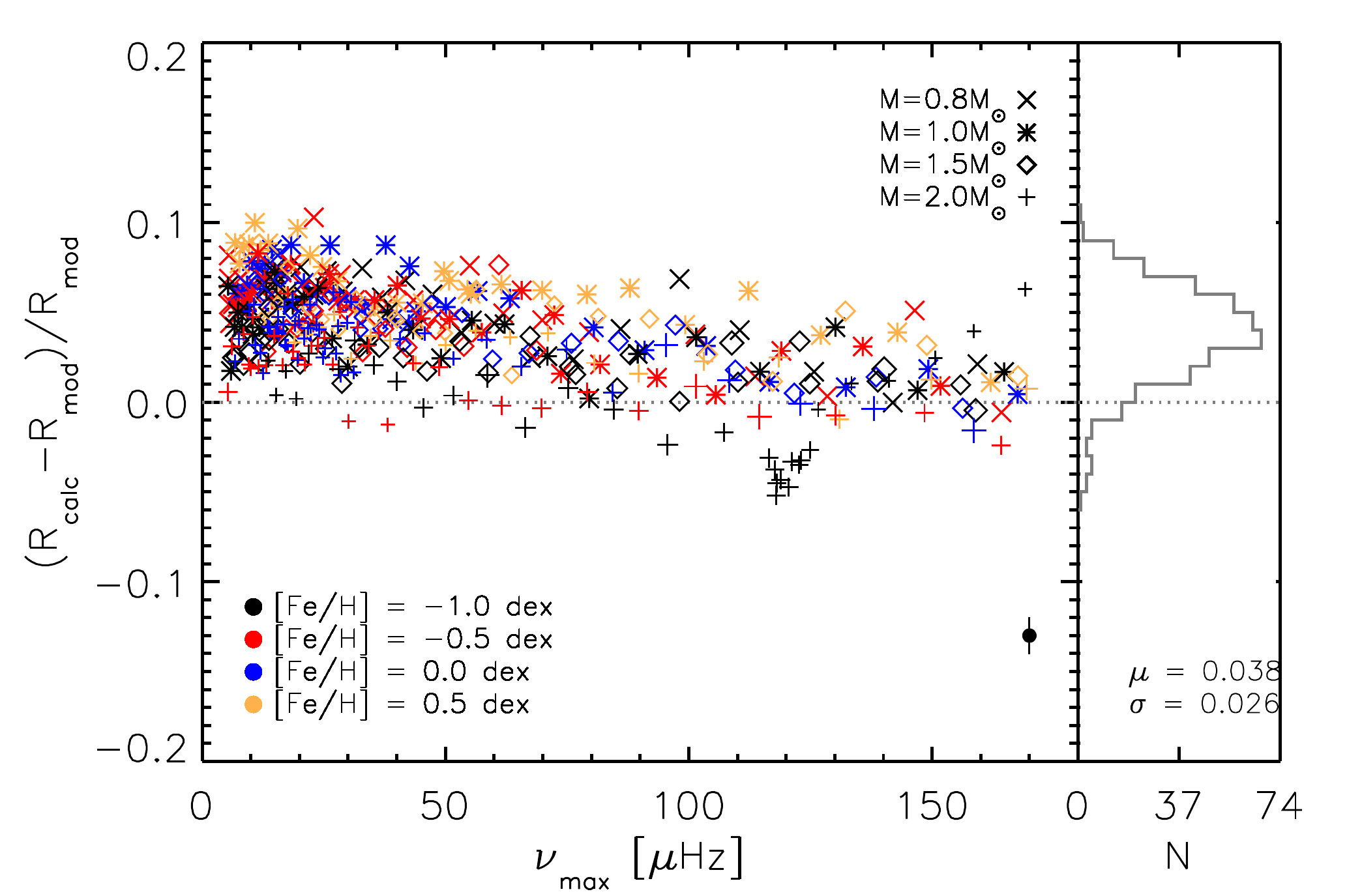}
\end{minipage}
\begin{minipage}{0.45\linewidth}
 \includegraphics[width=\linewidth]{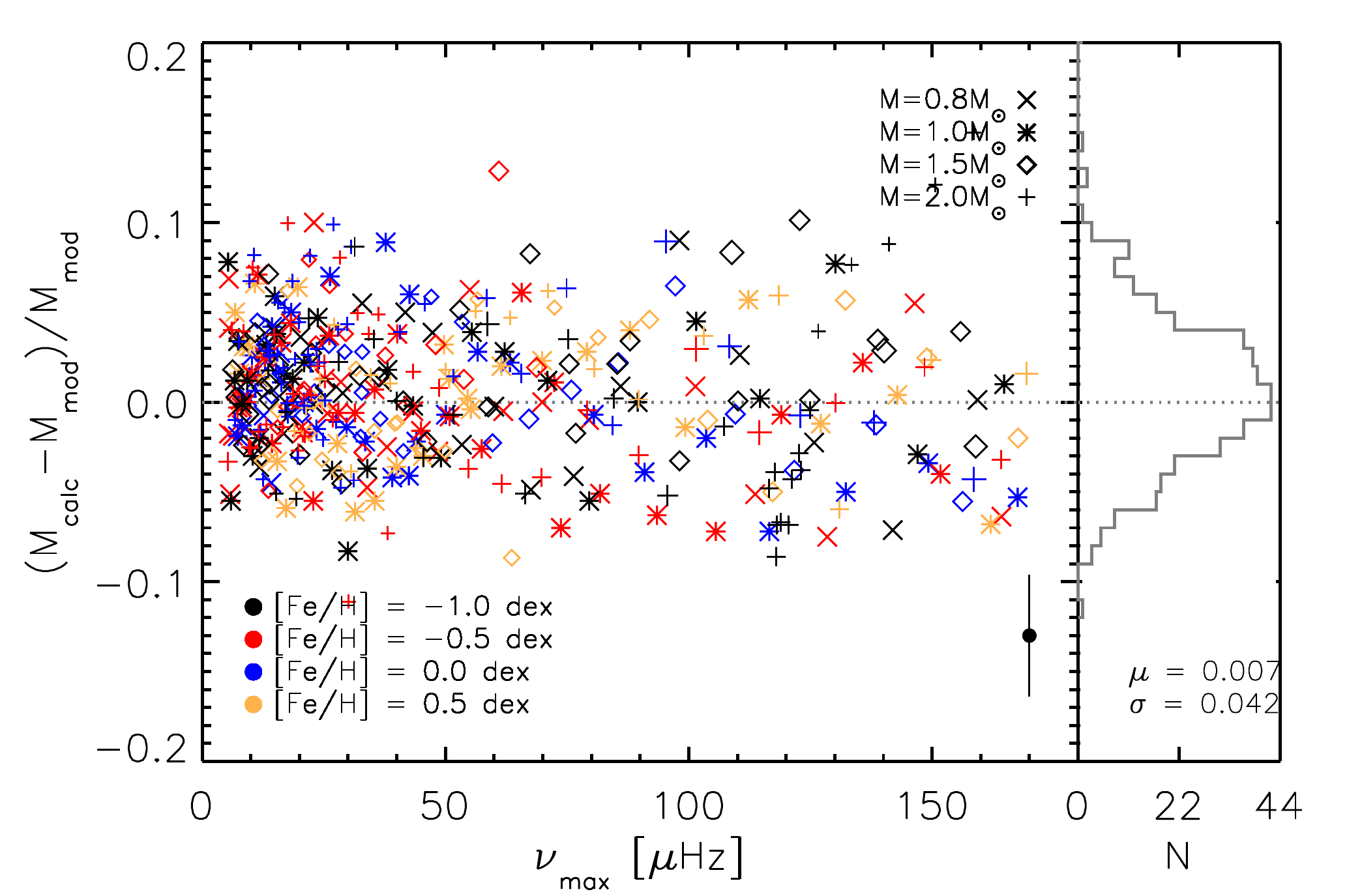}
\end{minipage}
 \begin{minipage}{0.45\linewidth}
 \includegraphics[width=\linewidth]{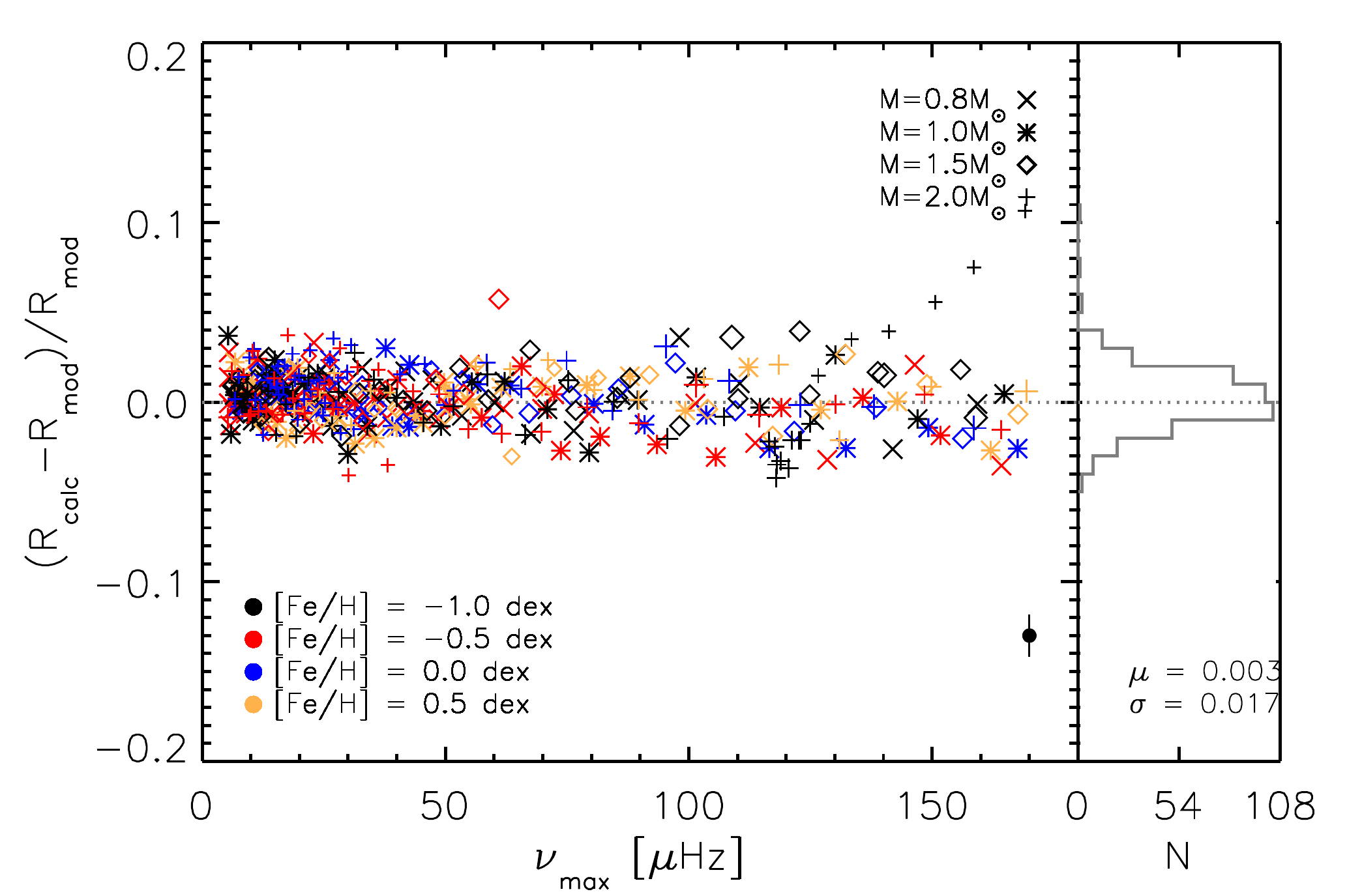}
\end{minipage}
 \caption{Same as Fig.~\ref{fig:MR} but now with the initial values of $T_{\rm eff}$, $\nu_{\rm max}$ and [Fe/H] perturbed with typical uncertainties (see text for more details)}
 \label{fig:MRperturb}
\end{figure*}



\begin{table}
	\centering
	\caption{Mean and standard deviation of the fractional differences in mass and radius between model values and values computed with different reference functions. The first column indicates the origin of the value used for the $\Delta\nu$ reference. Perturbed indicates that the initial values of the Monte Carlo were perturbed (see text for more detail).}
	\label{tab:res}
	\begin{tabular}{lrrrr} 
		\hline
method & $\mu_{\mathrm{mass}}$ & $\sigma_{\mathrm{mass}}$ & $\mu_{\mathrm{radius}}$ & $\sigma_{\mathrm{radius}}$ \\
\hline
solar & 0.075 & 0.050 & 0.037 & 0.024\\
Eq.~\ref{eq1} & 0.000 & 0.022 & 0.000 & 0.011\\
solar (perturbed) & 0.079 & 0.059 & 0.038 & 0.026\\
Eq.~\ref{eq1} (perturbed) & 0.007 & 0.042 & 0.003 & 0.017\\
Paper~I & 0.011 & 0.026 & 0.005 & 0.013\\
Eq.~\ref{eq2} & 0.001 & 0.014 & 0.001 & 0.007\\
Paper~I (perturbed) & 0.015 & 0.049 & 0.006 & 0.020\\
Eq.~\ref{eq2} (perturbed) & 0.007 & 0.045 & 0.003 & 0.018\\
		\hline
	\end{tabular}
\end{table}

\section{Results}
As mass is not known \emph{a priori}, the procedure we have developed is iterative in nature. 
An initial estimate of the input mass is determined using the classical \citep{kb95} scaling relation and the corresponding solar reference values of \dnu$_{\rm ref}$~=~\dnu$_\odot$~=~135.1~$\pm$~0.1~$\mu$Hz, \citep{hub11}, and $\nu_{\rm max,ref}$~=~\numax$_\odot$~=~3050~$\mu$Hz \citep{kb95}. This mass is then used in Eq.~\ref{eq1} to compute an improved mass. We iterate until the mass remains constant to a level of 0.001~M$_{\odot}$ which is typically reached in 2-5 iterations. 


We first demonstrate the efficacy of our fit by predicting the masses and radii for models used in the training grid. We produce 1000 instantiations of each `star'  perturbing each observable quantity with Gaussian noise according to a typical measurement uncertainty. We assume uncertainties of 5\% on $\nu_{\rm max}$, 0.09~dex in [Fe/H], 85~K in $T_{\rm eff}$ and 0.01~$\mu$Hz on the frequencies. In this test each faux observable quantity was perturbed around the known values from the stellar model. 
The relative differences in mass and radius between the values calculated using the scaling relations and the real values are shown in Fig.~\ref{fig:MR}. The mean and standard deviation indicated in each panel of Fig.~\ref{fig:MR} and in Table~\ref{tab:res} indeed show that with the new reference function (Eq.~\ref{eq1}) the biases in both mass and radius are mitigated and that the standard deviation is reduced by a factor of 2. Additionally, the dependence on $\nu_{\rm max}$ is mitigated by the reference function.




In a second approach we computed an initial mass using the reference function described in Paper~I. This mass is then used as an input to the reference function (Eq.~\ref{eq2}) to iteratively compute the mass and radius. As per the first function we evaluate the function by perturbing the model data. The results are shown in Figure~\ref{fig:MR_a}. Again with this function we mitigate the bias of the relative difference in calculated and model values and reduce the standard deviation by roughly a factor of 2. However, this results in standard deviations that are smaller (about $2/3$) compared to the results from the first approach.



In addition to the test above we also used used perturbed values as our central values. We perturbed the mass, metallicity, effective temperature and individual frequencies in the same manner as before and computed masses and radii using the scaling relations iteratively with the reference functions described in Eqs~\ref{eq1} and \ref{eq2}. The results are shown in Figs~\ref{fig:MRperturb} and \ref{fig:MRperturb_a}. These figures show that the reference functions that we present here indeed remove trends that are present in the mass and radius determinations as a function of $\nu_{\rm max}$ and mitigate the biases to a large extent. However, the standard deviations from using either Eq.~\ref{eq1} or Eq.~\ref{eq2} are very similar and significantly larger than in the `ideal' case, showing the impact of uncertainties in the observations.

\subsection{Hare-and-Hound Validation}
\begin{figure*}
\centering
\includegraphics[width=\textwidth]{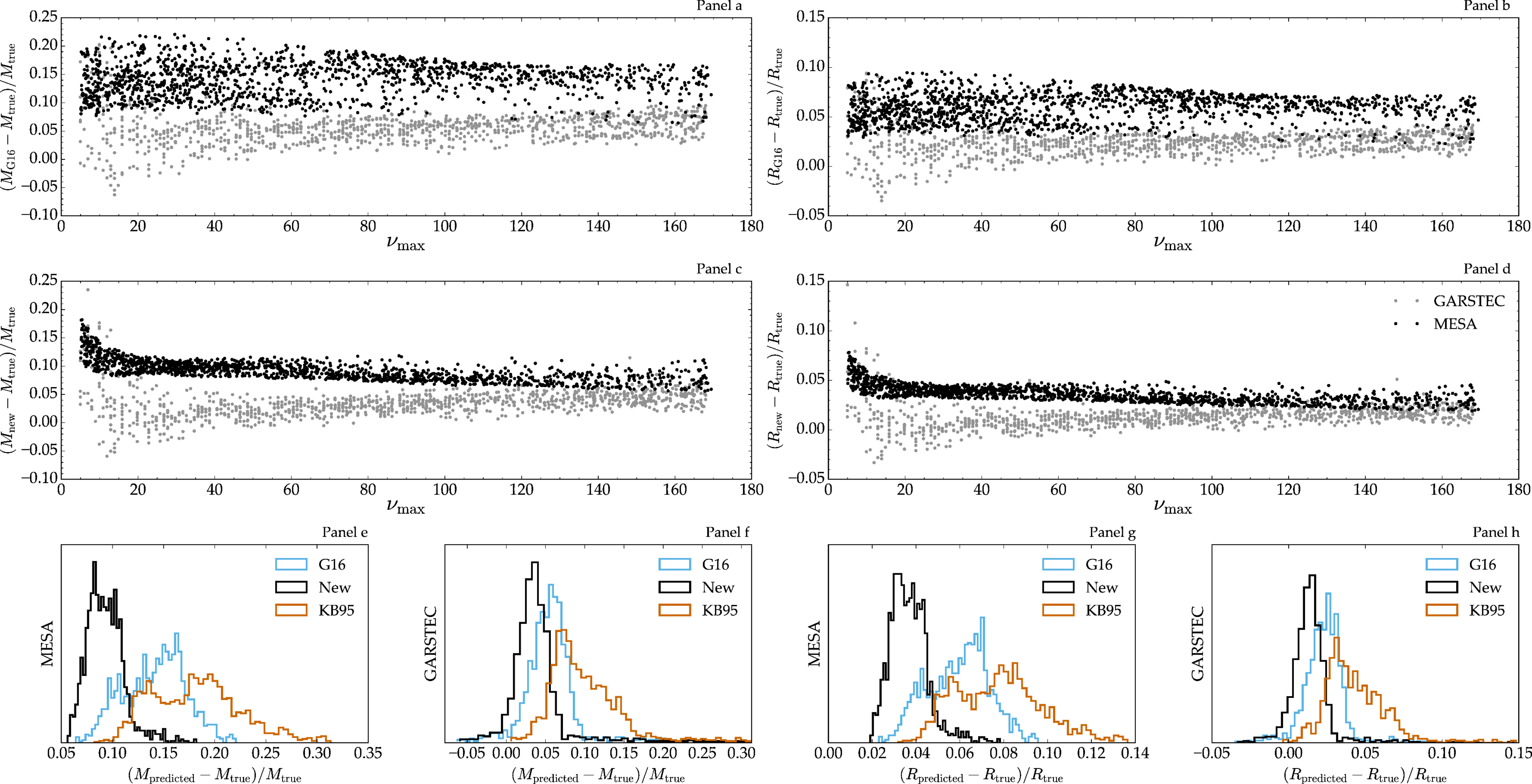}
\caption{Results from the hare-and-hound validation for mass (left) and radius (right) using MESA (black) and GARSTEC models (grey). The top panels show results using reference values from Paper~I and the central panels show results using Eq.~\ref{eq1}. The histograms show the distributions of the results for the solar reference (orange, KB95), the reference from Paper~I (blue, G16) and Eq.~\ref{eq1} for the codes indicated on the y-axis.}
\label{fig:HH}
\end{figure*}

We have thus far demonstrated that the new functions improve mass and radius estimations for the data they fit.
In order to prove the general efficacy of our new corrections we appraise the performance of the functions on data to which they have not been exposed. 
We perform a series of hare-and-hound validations using stellar models calculated with two additional evolution codes. 
We used GARSTEC \citep{weiss08} and MESA \citep{pax11} to compute grids of stellar tracks from which we draw testing data. 
Over 1000 models from each grid were extracted to serve as hares with the requirements that each selected model falls within the verified mass,  metallicity and \numax\ ranges of the correction functions.
The choice of microphysics and modelling assumptions differ between these two codes and they also differ from the code used to calculate the training data in YREC. 
Thus mass and radius predictions of the hares will indicate the degree of stellar evolution code dependence in our function. 

Both the GARSTEC and MESA grid were computed using  OPAL opacities \citep{ir96} in conjunction with \citet{ferg05} tables at low temperatures. We utilized the updated 2005 tables for the OPAL equation of state \citep{rog02} along with the assumption of an Eddington plane-parallel atmosphere. We have taken the recommended  nuclear reaction rates from the NACRE collaboration in combination with those from \citet{CF88}. The treatment of convection follows the mixing length theory as described by \citet{KWW12} and models were computed with the respective solar calibrated mixing lengths. Radial mode frequencies were calculated with the ADIPLS pulsation package \citep{jcd2008} for the GARSTEC models and with the GYRE pulsation package \citep{townsend2013} for the MESA models. The large frequency separation determined using a weighted median around the scaled \numax\ as per \citet{BA2016}. 

The grids differ in that the GARSTEC models assume \citet{Asp09} solar abundances and were computed without overshoot or atomic diffusion. In contrast, the MESA grid employs the \citet{GS98} solar composition.
For the MESA models we use a step overshooting prescription at all convective boundaries with an efficiency of 0.08 pressure scale heights. Additionally, atomic diffusion as per \citet{Thoul94} is included in tracks  up to  $M \le  1.2\, \rm{M}_{\odot}$ and its efficiency smoothly reduced thereafter. 
 
The hares comprise 1095 models from the GARSTEC grid with 15 models extracted along the RGB of 73 different evolutionary tracks. We also include 1492 models from the MESA grid having selected four models from 373 individual tracks.

The results of the hare-and-hound validation for the extracted models are presented in the seven panels in  Fig.~\ref{fig:HH}. 
Figures~\ref{fig:HH}a~and~\ref{fig:HH}b show the relative errors in the predictions for mass and radii using the correction outlined in Paper~I. 
The corresponding relative differences after applying  Eq.~\ref{eq1} are shown in Figs~\ref{fig:HH}c~and~\ref{fig:HH}d. 
In Figures~\ref{fig:HH}e-h we plot the error distributions for each code and parameter. 
We include in these panels the predictions from the Paper~I correction (G16), Eq.~\ref{eq1} (New) as well as the classic \citet{kb95} scaling relation (KB95). 

Our results indicate that there is a systematic offset between MESA models and those from the other two codes. The offset can be explained by a different temperature scale in the RGB models that is caused by a larger mixing-length parameter used in the MESA models that originates from a solar calibration that includes elemental diffusion. The corresponding change to \numax\ and $T_{\rm eff}$ by reducing the MESA models by 15\,K is sufficient to centre its error distributions around the zero point (such that they lay on top of the GARSTEC points). We note also that the GARSTEC models diverge from the YREC models as a function of evolution as seen by the trend in \numax\ which is to be expected when comparing models with and without atomic diffusion (see e.g. \citealt{2004ApJ...606..452M}). 

\section{Discussion and Conclusions}
We have found two reference functions for the scaling relation of the large frequency separation that mitigate the mass dependence. We have shown that these results are valid for stellar models in the range 5~$\mu$Hz~$ <  \nu_{\rm max} < $~170~$\mu$Hz and improve the mass and radius determinations by 10\% and 5\% respectively (compared to the solar reference). This is true in the limit of ideal data obtained from canonical stellar models and no surface effects in which these functions provide a better underlying model to use when applying the $\Delta\nu$ scaling relation. Therefore, this method can be assumed to be valid for stars that do not show any extreme physics (e.g. large magnetic fields).

If we assume observational uncertainties on $T_{\rm eff}$, $\nu_{\rm max}$ and [Fe/H] the devised reference functions continue to improve the accuracy of the masses and radii as in the ideal case. However, the precision depends on the observational uncertainties. In theory it is better to fit the residuals (hence the equation from Paper~I combined with Eq.~\ref{eq2}). However, in practice observational uncertainties dominate and there is minor difference between using Eq.~\ref{eq1} and Eq.~\ref{eq2}. 

We have also tested the reference functions on a large number of models computed with MESA \citep{pax11} and GARSTEC \citep{weiss08}. From these tests it is clear that the underlying form is correct. However, we found that a slight offset in temperature will propagate to differences in $\nu_{\rm max}$ that in turn influence the accuracy of the derived masses and radii. Optimising the coefficients of the functions described in Eqs~\ref{eq1} and \ref{eq2} would mitigate this offset. 

We note more generally, that our reference functions are particularly sensitive to variations in $\nu_{\rm max}$. Therefore, a change in the computation of $\nu_{\rm max}$ \citep[see e.g.][]{viani2017}, or the temperature scale of the models will require a recalibration of the coefficients (see Table~\ref{tab:param}).

\section*{Acknowledgements}

The research leading to the presented results has received funding from the
European Research Council under the European Community's Seventh Framework
Programme (FP7/2007-2013) / ERC grant agreement no 338251 (StellarAges).
S.B. acknowledges partial support of NASA grant NNX16AI09G and NSF grant AST-1514676.





\appendix

\section{Fits and Residuals}
Here we present the figures obtained using first the reference function presented in Paper~I and subsequently the fits as described in Eq.~\ref{eq2}.

\begin{figure*}
\centering
 \begin{minipage}{0.45\linewidth}
 \includegraphics[width=\linewidth]{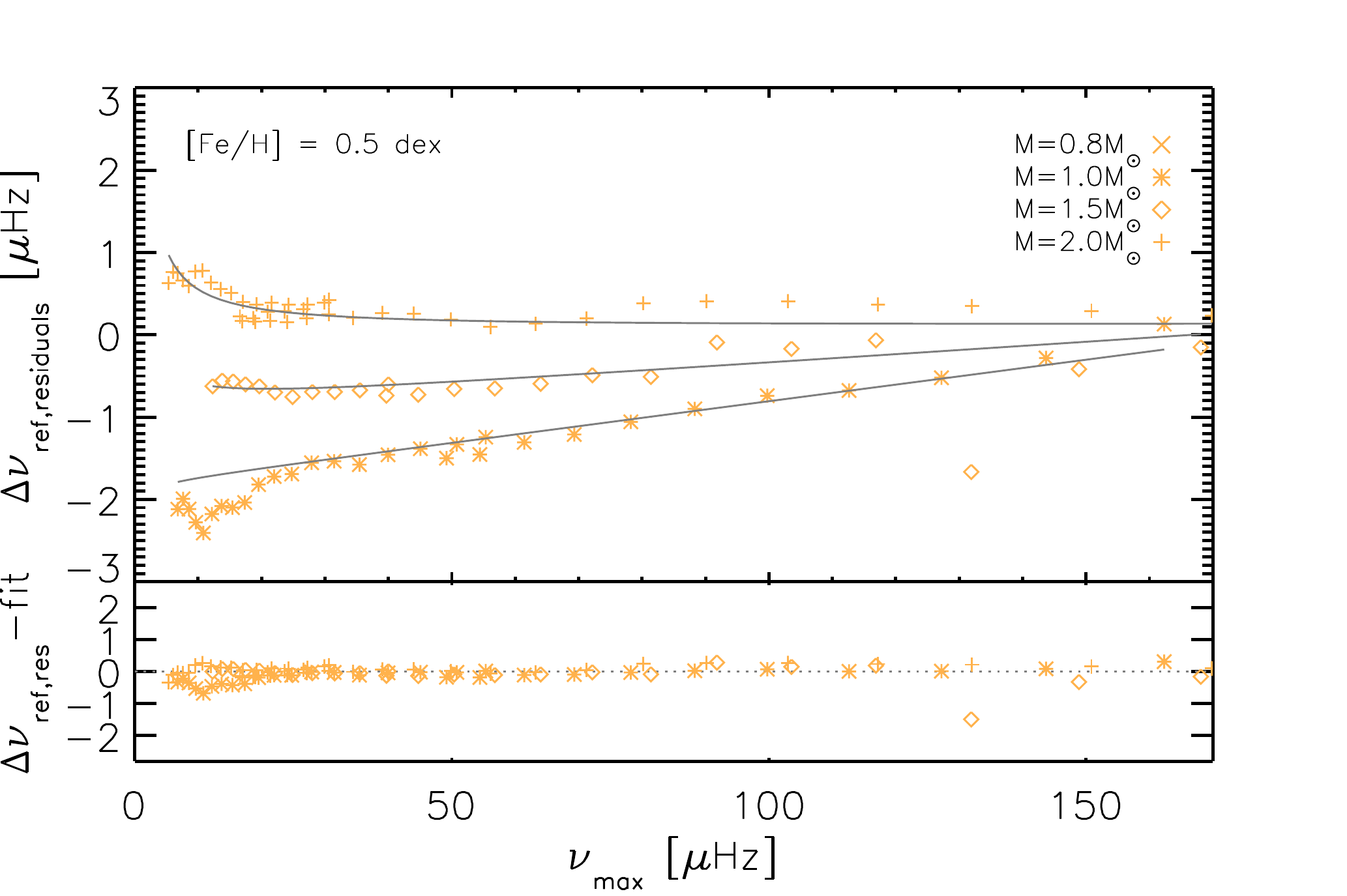}
\end{minipage}
 \begin{minipage}{0.45\linewidth}
 \includegraphics[width=\linewidth]{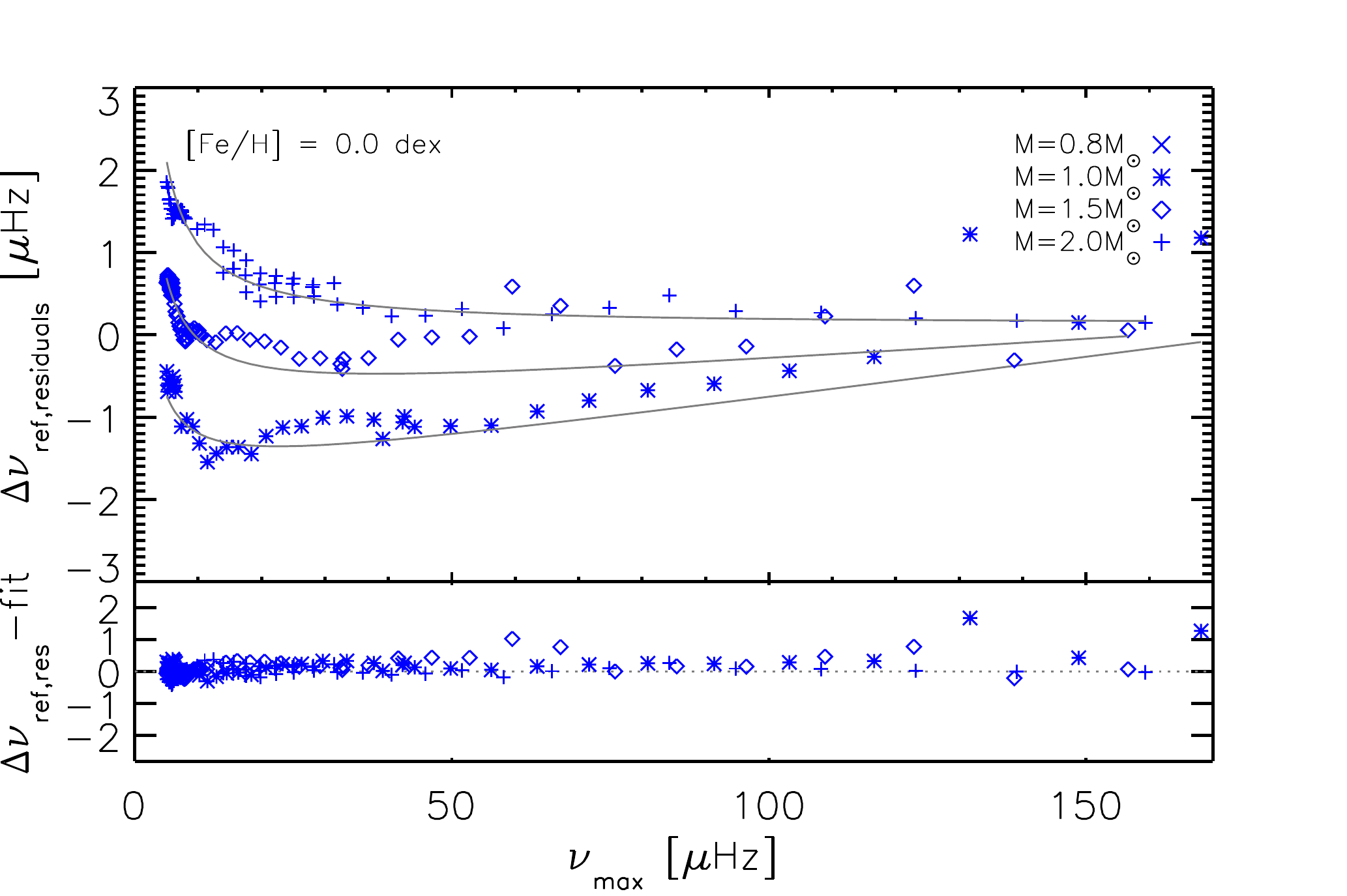}
\end{minipage}
 \begin{minipage}{0.45\linewidth}
 \includegraphics[width=\linewidth]{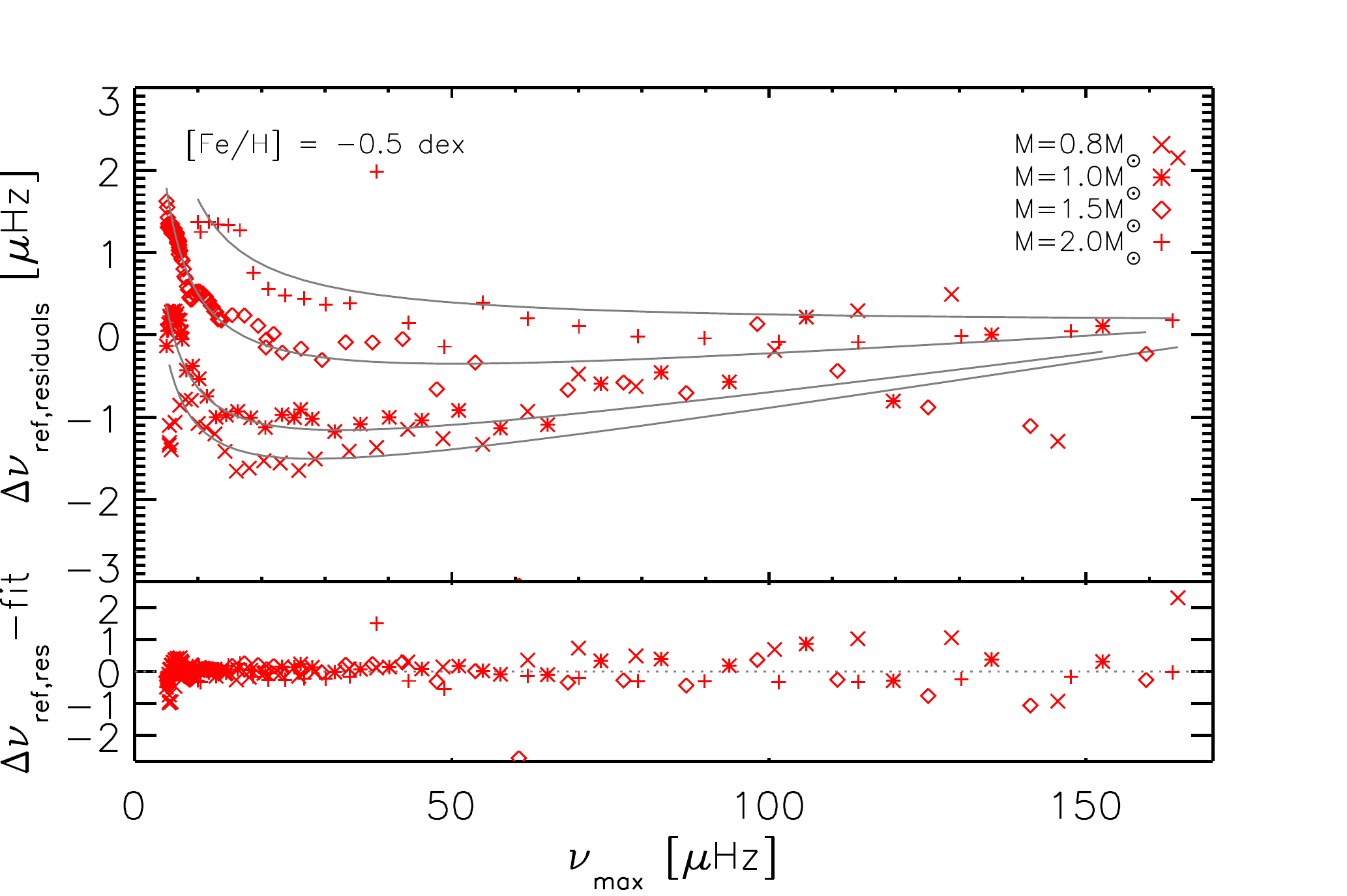}
\end{minipage}
 \begin{minipage}{0.45\linewidth}
 \includegraphics[width=\linewidth]{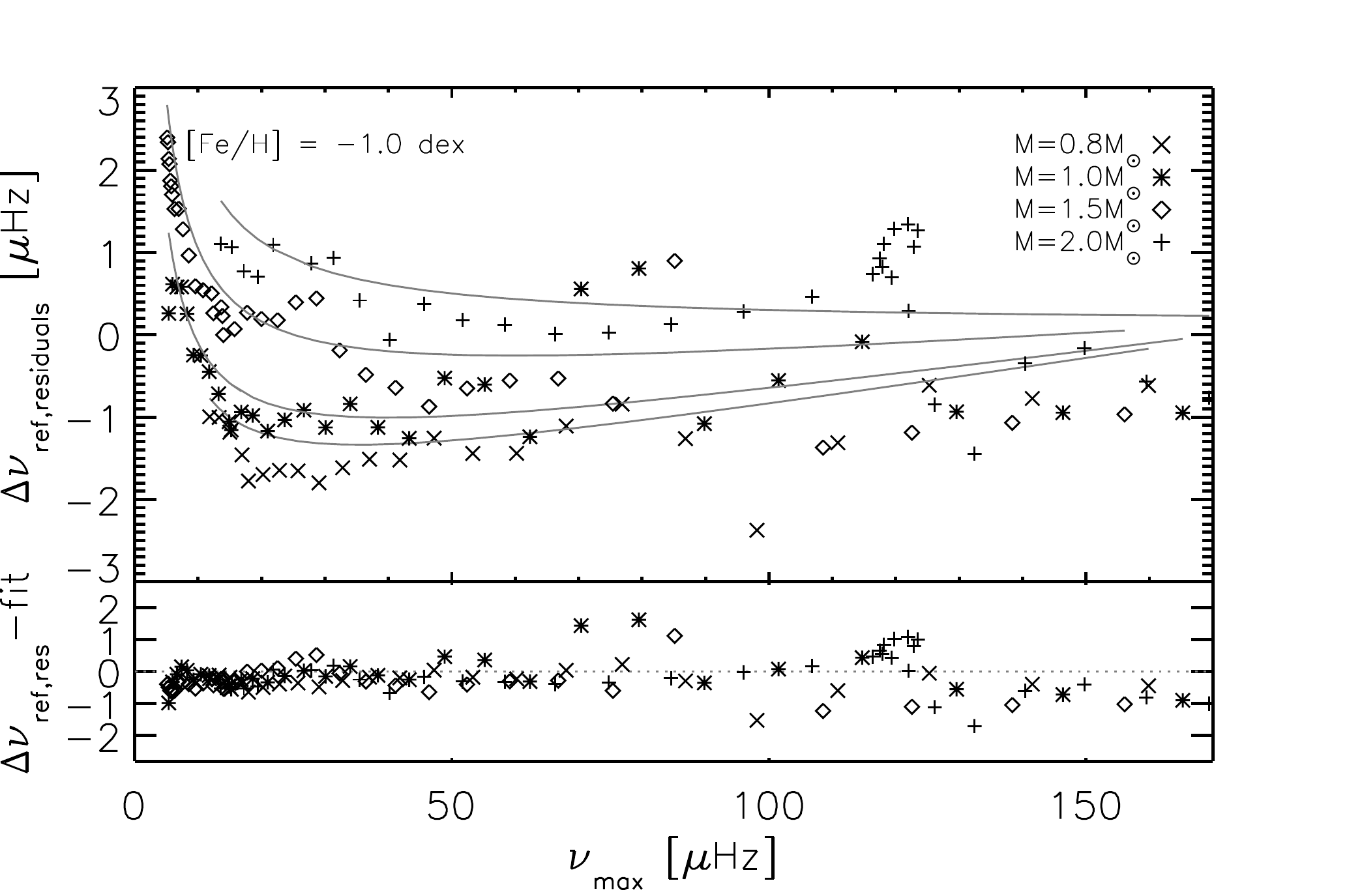}
\end{minipage}
 \caption{Residuals of the reference values for \dnu\ obtained from the equation in Paper~I as a function of \numax\ with in grey the mass dependent fit (Eq.~\ref{eq2}). Each panel shows one metallicity and different masses indicated in the legend. Residuals between the data points and the fit are shown in the lower part of each panel with the grey dotted line indicating zero.}
 \label{fig:fit_a}
\end{figure*}

\begin{figure*}
\centering
\begin{minipage}{0.45\linewidth}
 \includegraphics[width=\linewidth]{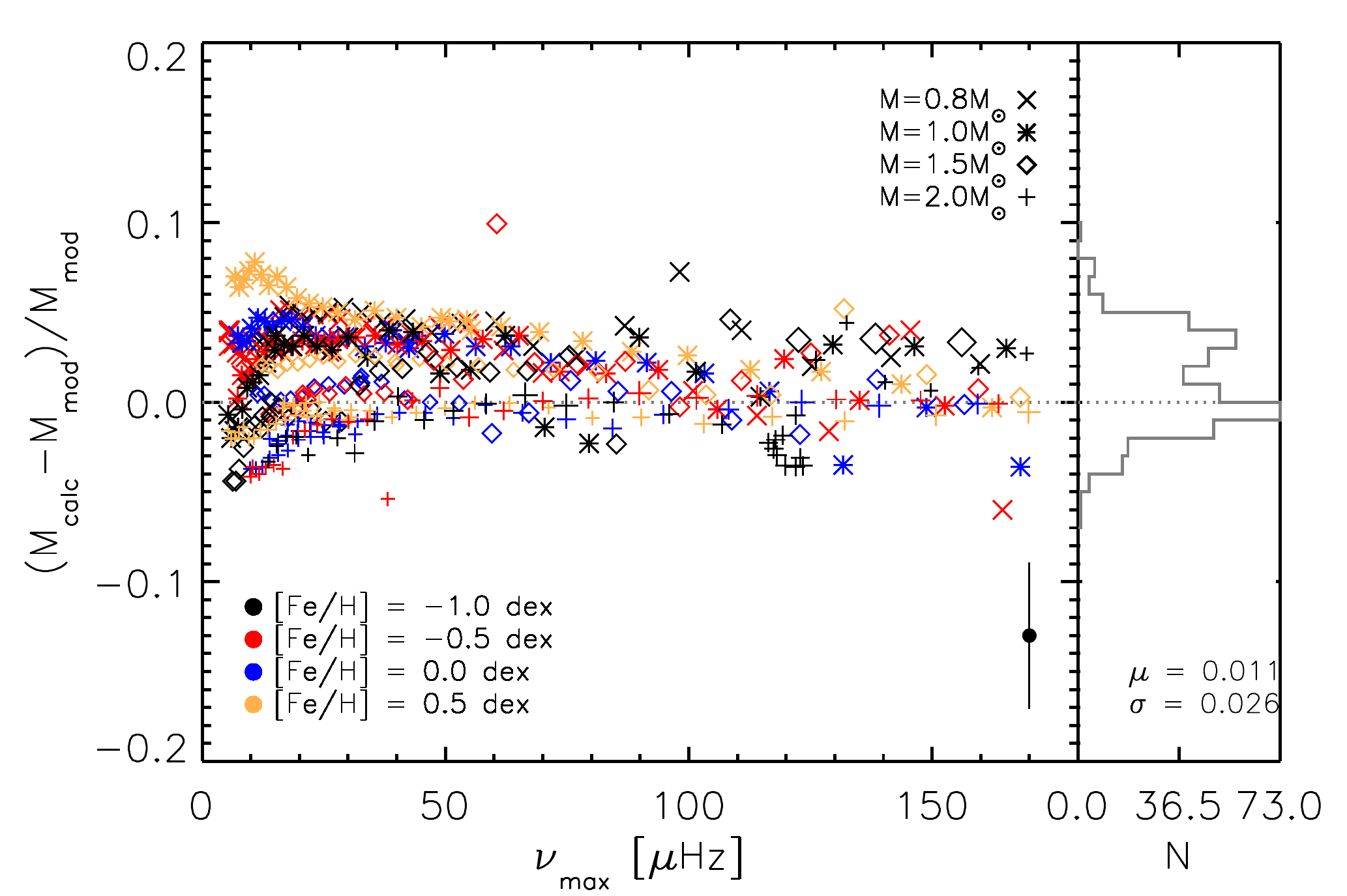}
\end{minipage}
 \begin{minipage}{0.45\linewidth}
 \includegraphics[width=\linewidth]{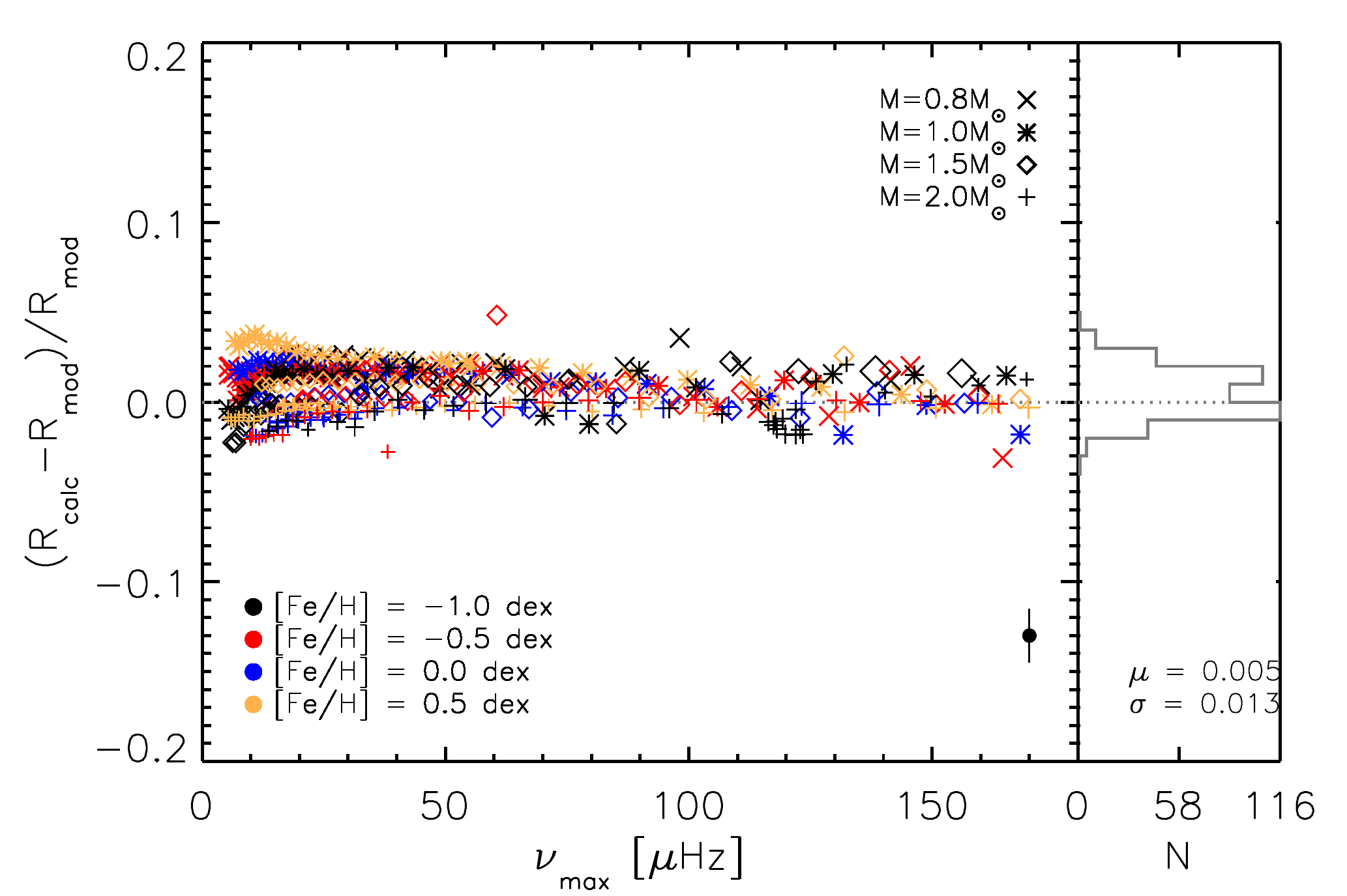}
\end{minipage}
\begin{minipage}{0.45\linewidth}
 \includegraphics[width=\linewidth]{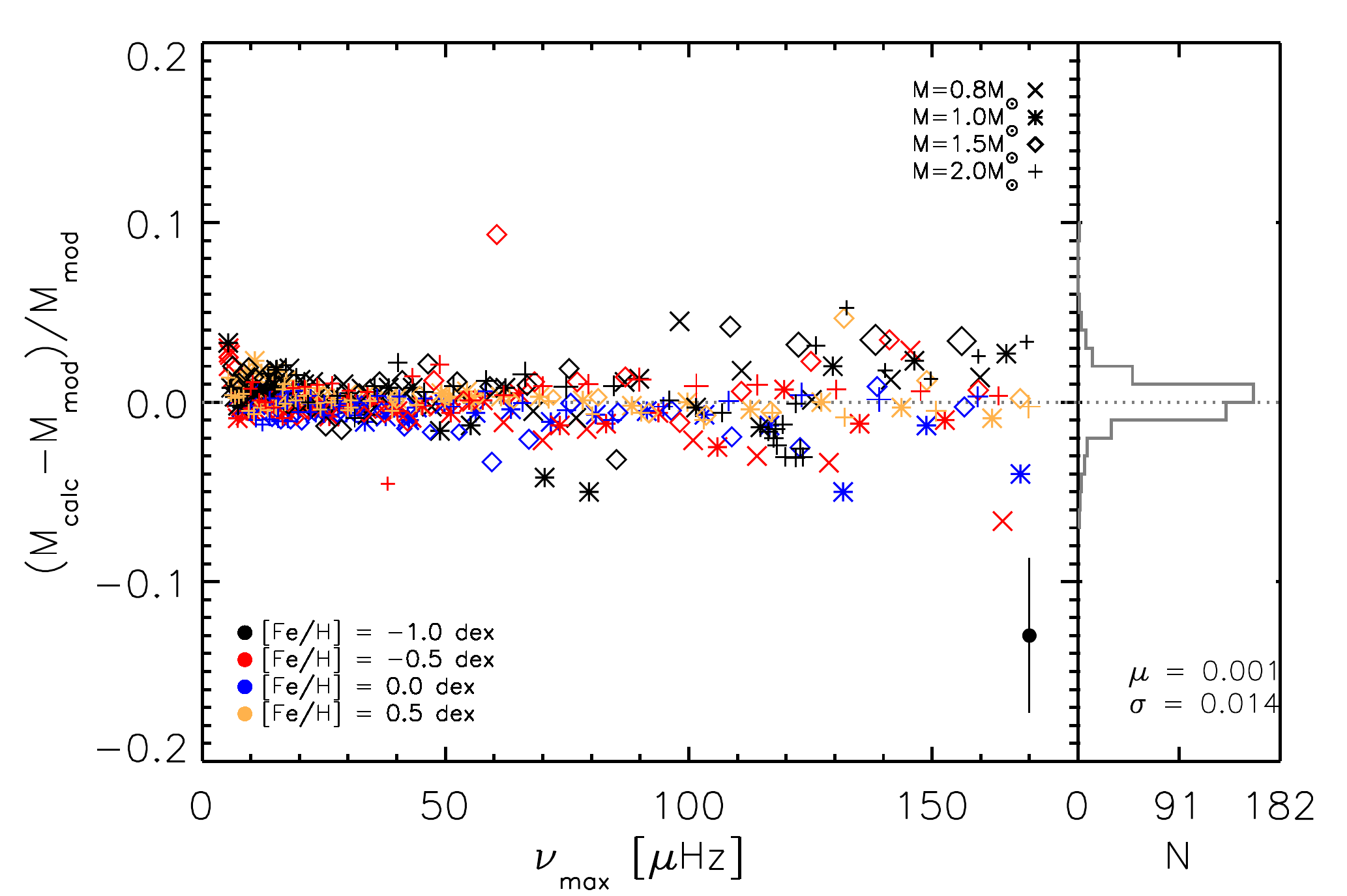}
\end{minipage}
 \begin{minipage}{0.45\linewidth}
 \includegraphics[width=\linewidth]{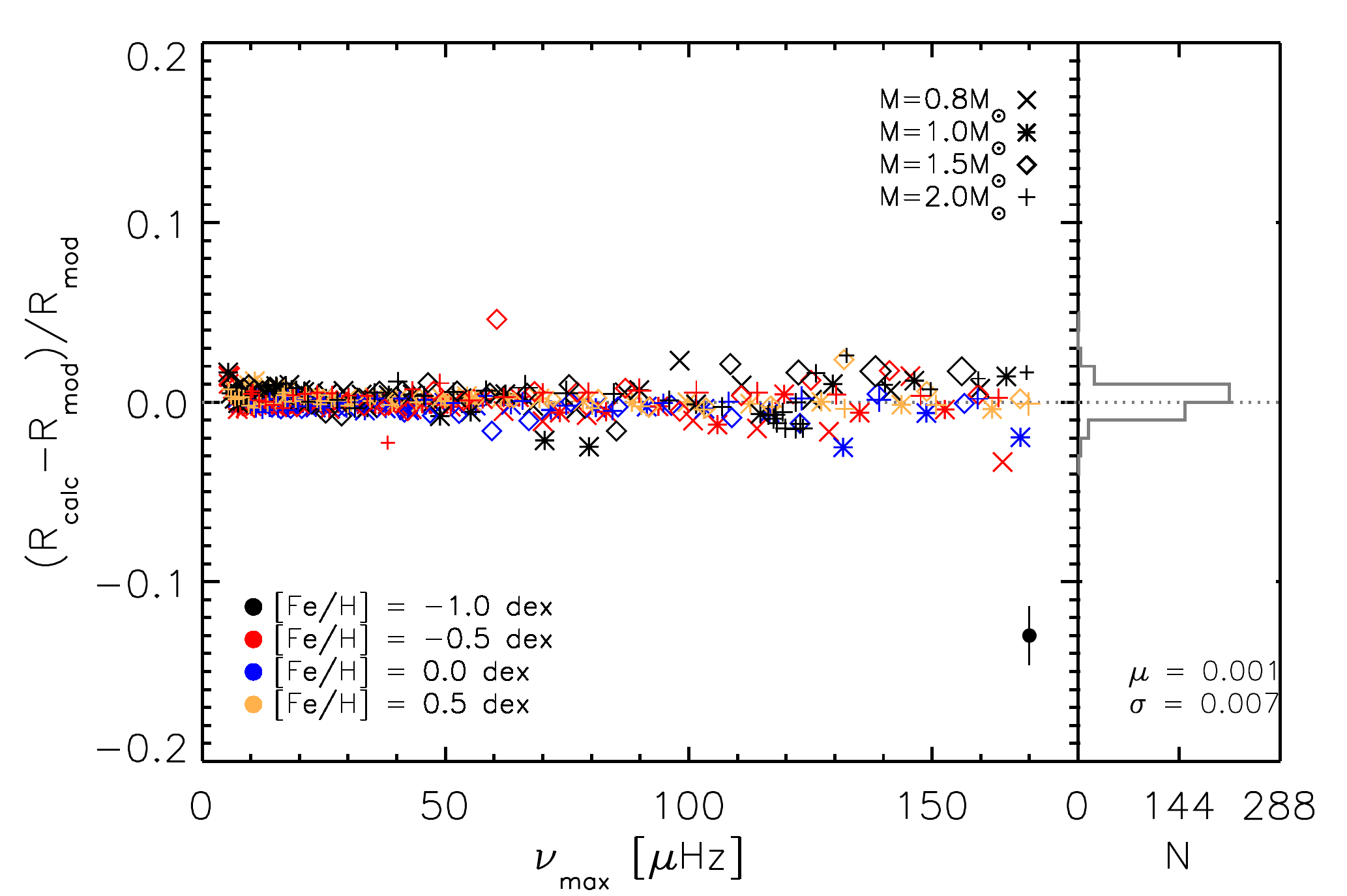}
\end{minipage}
\caption{Same as Fig.~\ref{fig:MR} but now using the function presented in Paper~I (top) and Eq.~\ref{eq2} (bottom) as a reference for \dnu\ to obtain the mass (left) and radius (right).}
\label{fig:MR_a}
\end{figure*}

\begin{figure*}
\centering
\begin{minipage}{0.45\linewidth}
 \includegraphics[width=\linewidth]{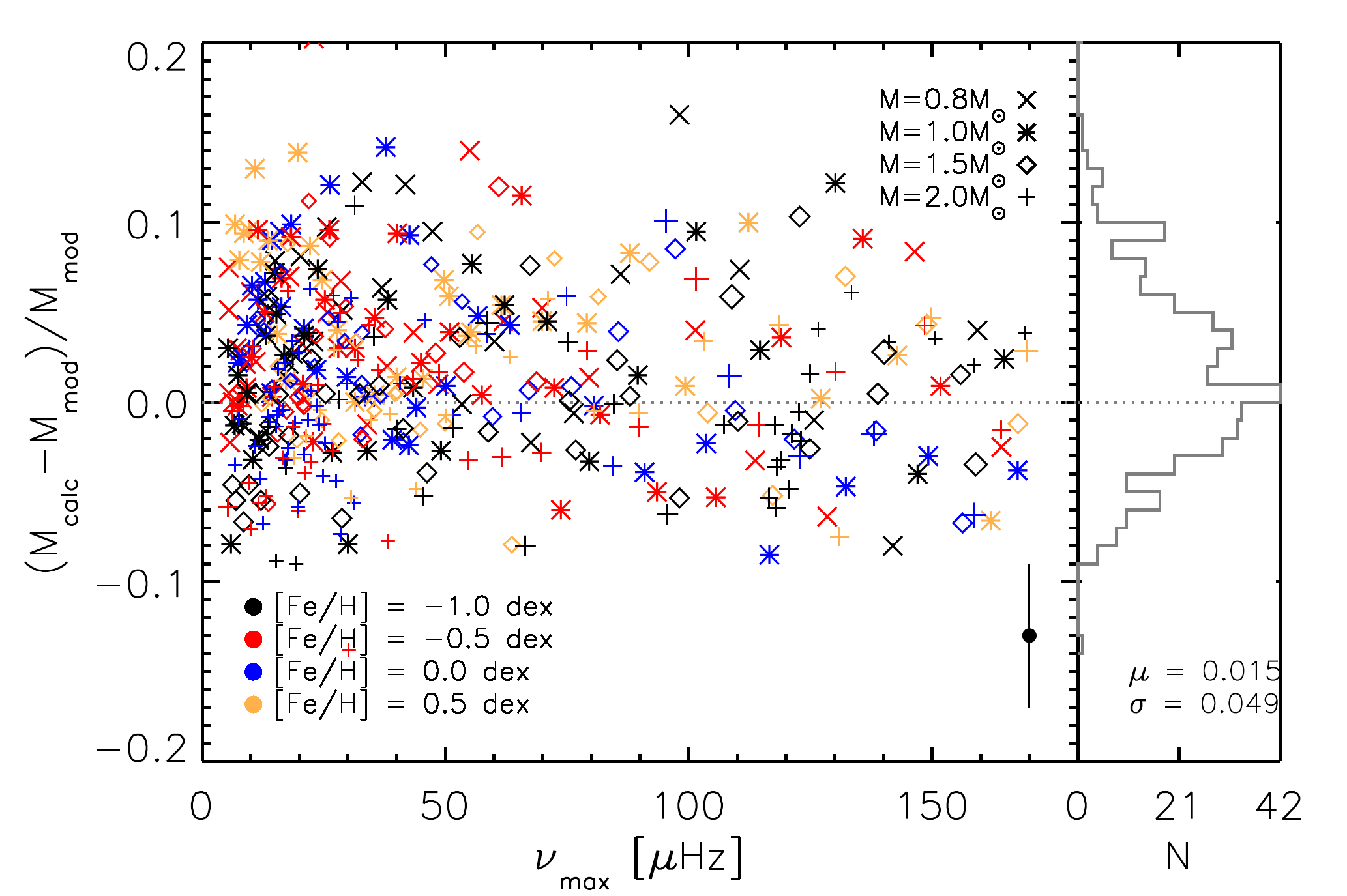}
\end{minipage}
 \begin{minipage}{0.45\linewidth}
 \includegraphics[width=\linewidth]{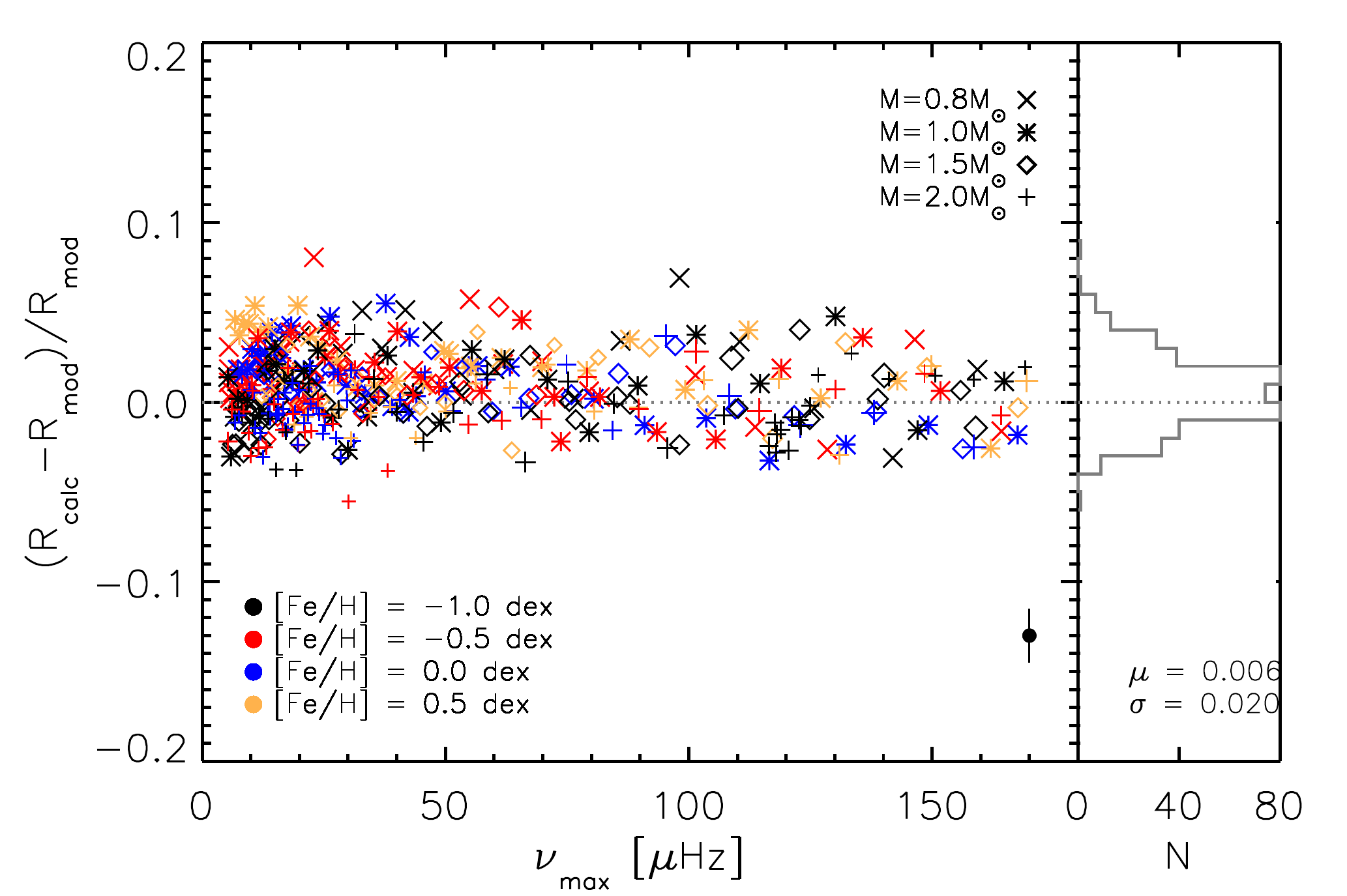}
\end{minipage}
\begin{minipage}{0.45\linewidth}
 \includegraphics[width=\linewidth]{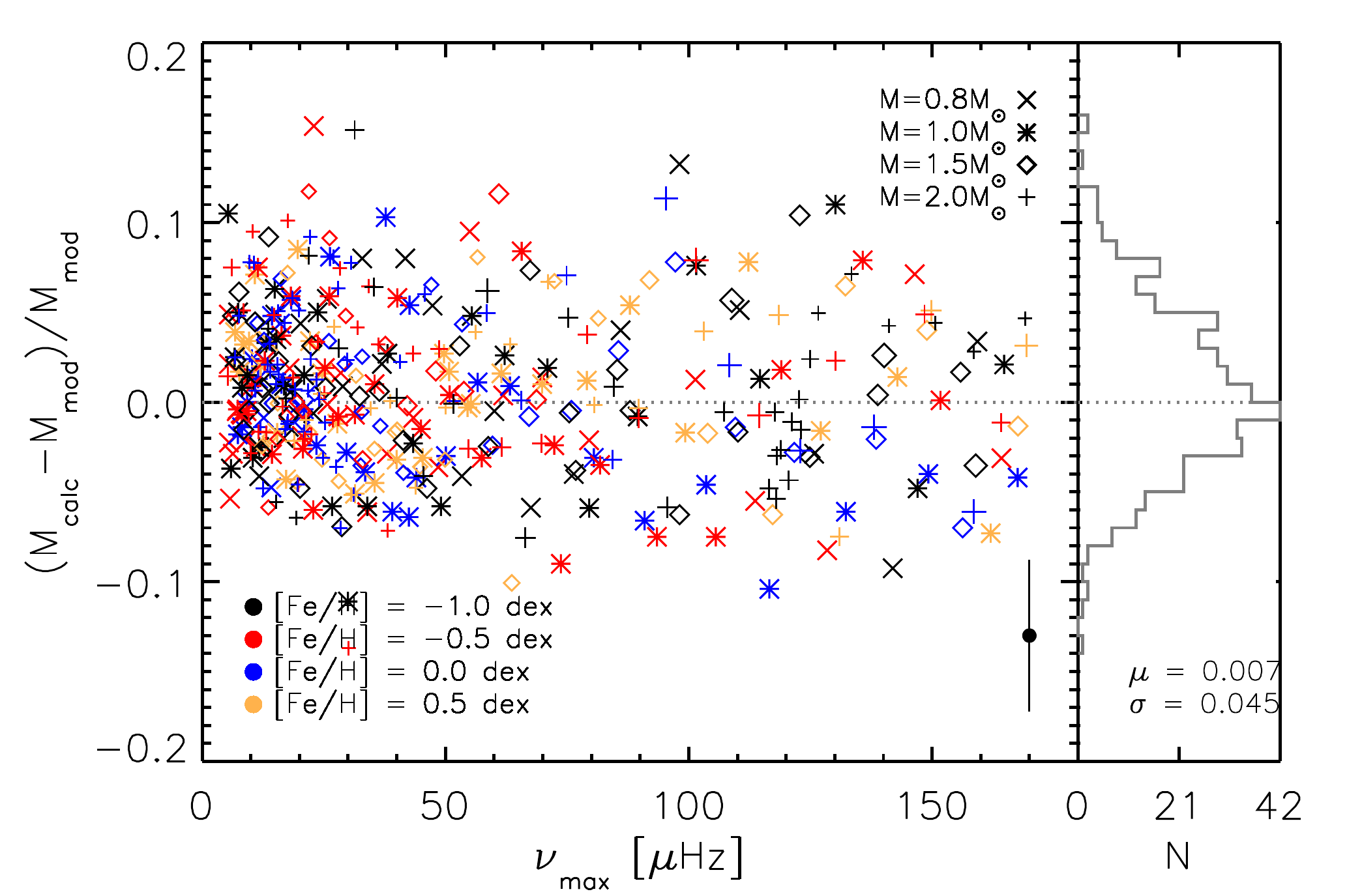}
\end{minipage}
 \begin{minipage}{0.45\linewidth}
 \includegraphics[width=\linewidth]{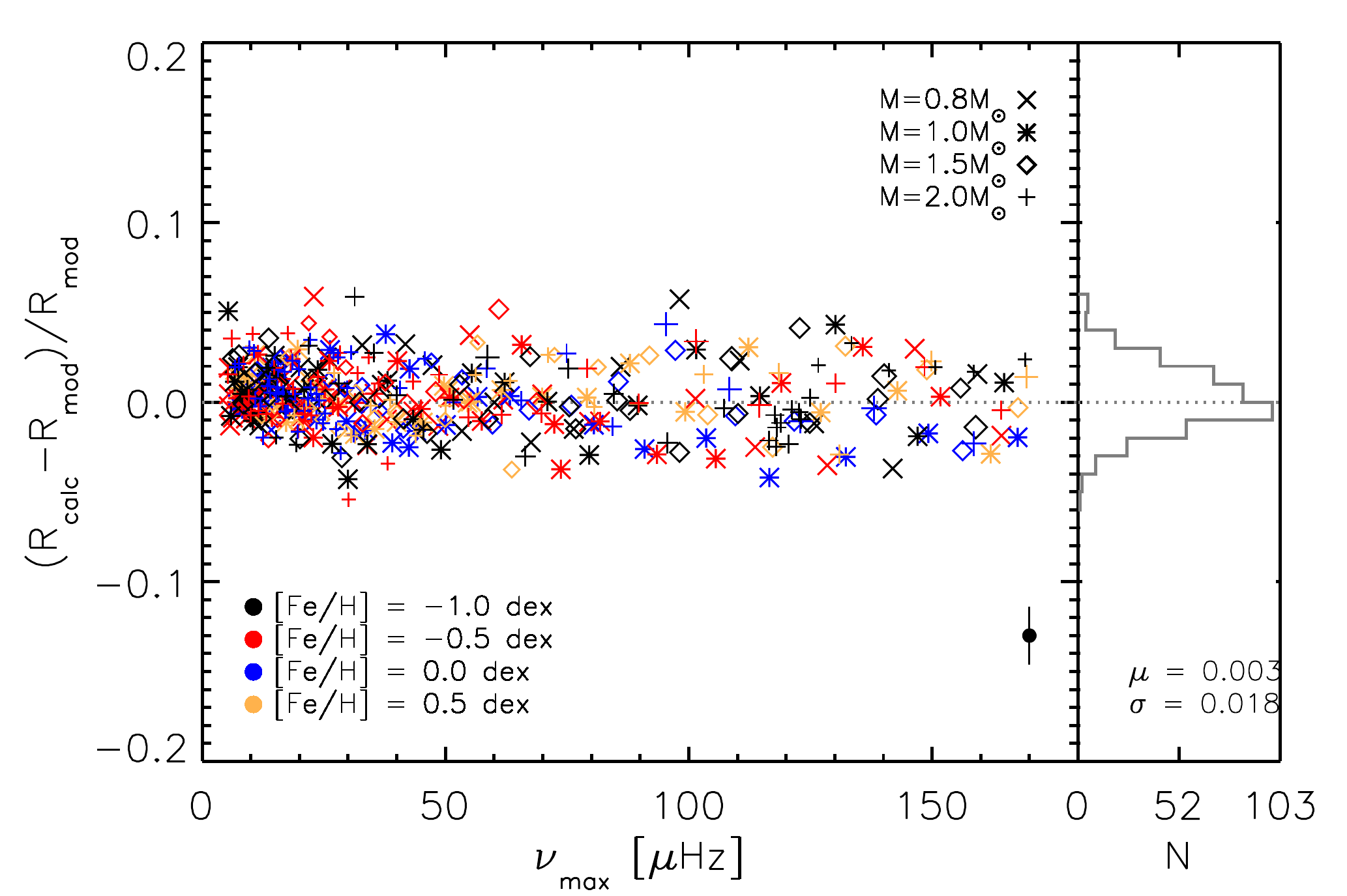}
\end{minipage}
 \caption{Same as Fig.~\ref{fig:MRperturb} but now for the \dnu\ reference computed using the equation in Paper~I (top) and Eq.~\ref{eq2}.}
 \label{fig:MRperturb_a}
\end{figure*}

\clearpage
\onecolumn

\section{Example Code}

Here we provide some example code in both Python and IDL of the iterative process to determine masses and radii with the reference function for $\Delta\nu$ presented in this work.

\subsection{Python}
\begin{lstlisting}[language=Python]
def apply_corr(inst):
   """A function that corrects the $\Delta_nu$ reference used in the scaling relations and predicts massed and radii.  Written for clarity rather than efficiency.

      inst -- A list of observables required by the scaling relations [Teff, Nu_Max, [Fe/H], Dnu]
      pred -- A list of predictions from the different scaling relations """
      
   import math 
      
   #Function coefficients as per Table 1   
   A1=124.7236
   A2=2.23354
   A3=17.61178
   A4=0.73025
   A5=0.01981
   A6=0.93173

   B1=1.88
   B2=0.02
   B3=5.14
   B4=10.90
   B5=3.69
   B6=0.01
 
 
   #Solar values of observables   
   DNU_Sun=135.1
   Nu_max_Sun=3050.0
   Teff_Sun=5772.0

   #Unpack list of observables          
   teff_obs=inst[0]
   nu_max_obs=inst[1]
   feh_obs=inst[2]
   dnu_obs=inst[3]
   pred=[]

   #Classic Scaling Relation --  Kjeldsen & Bedding (1995)
   KB95_mass = (nu_max_obs/Nu_max_Sun)**3.*(dnu_obs/DNU_Sun)**(-4.)*(teff_obs/Teff_Sun)**(3./2.)
   KB95_rad = (nu_max_obs/Nu_max_Sun)*(dnu_obs/DNU_Sun)**(-2)*(teff_obs/Teff_Sun)**0.5
   pred +=[KB95_mass,KB95_rad]

   #Teff and metallicity correction -- Guggenberger et al. (2016)   
   G16_ref=(0.64*feh_obs+1.78)*math.exp(teff_obs*(-0.55*feh_obs+1.23)/10000.)*(math.cos(22.21*teff_obs/10000.+(0.48*feh_obs+0.12)))+0.66*feh_obs+134.92
   G16_mass=(nu_max_obs/Nu_max_Sun)**3.*(dnu_obs/G16_ref)**(-4.)*(teff_obs/Teff_Sun)**(3./2.)
   G16_rad=(nu_max_obs/Nu_max_Sun)*(dnu_obs/G16_ref)**(-2.)*(teff_obs/Teff_Sun)**0.5


   #Corrections presented in the current work 
   for i in range(5):
      #New Fit  -- First mass estimate from KB95 function 
      if i==0:
         G17New_mass=KB95_mass
      G17New_ref=A1+A2*G17New_mass+A3/nu_max_obs+A4*math.sqrt(nu_max_obs)-A5*nu_max_obs-A6*feh_obs

      #Fit from the residuals of G16 -- First mass estimate is from that function   
      if i==0:
         G17res_mass=G16_mass            
      G17res_ref=B1*G17res_mass+B2*nu_max_obs+(B3*G17res_mass-B4*feh_obs)/nu_max_obs-B5-B6* G17res_mass*nu_max_obs
      
      
      G17New_mass=(nu_max_obs/Nu_max_Sun)**3.*(dnu_obs/G17New_ref)**(-4.)*(teff_obs/Teff_Sun)**(3./2.)
      G17res_mass=(nu_max_obs/Nu_max_Sun)**3.*(dnu_obs/(G17res_ref+G16_ref))**(-4.)*(teff_obs/Teff_Sun)**(3./2.)

   G17New_rad=(nu_max_obs/Nu_max_Sun)*(dnu_obs/G17New_ref)**(-2)*(teff_obs/Teff_Sun)**0.5
   G17res_rad=(nu_max_obs/Nu_max_Sun)*(dnu_obs/(G17res_ref+G16_ref))**(-2)*(teff_obs/Teff_Sun)**0.5

   pred +=[G16_mass,G16_rad,G17New_mass,G17New_rad, G17res_mass, G17res_rad]
   return pred
\end{lstlisting}

\subsection{IDL}
\begin{lstlisting}[language=IDL]
  PRO computeMR,numax,dnu,Teff,feh,mcalc_org,mcalc,mres_org,mresc,rcalc_org,rcalc,rres_org,rresc
  ;call program with values for numax, dnu, Teff and feh
  ;returning masses (mcalc_org,mcalc,mres_org,mresc)
  ;and returning radii (rcalc_org,rcalc,rres_org,rresc)
  
  ;Function coefficients as per Table 1   
   A1=124.7236
   A2=2.23354
   A3=17.61178
   A4=0.73025
   A5=0.01981
   A6=0.93173

   B1=1.88
   B2=0.02
   B3=5.14
   B4=10.90
   B5=3.69
   B6=0.01
  
  ; iterative computation of mass with Eq. 3
   mcalc=(numax/3050.)^3.*(dnu/135.1)^(-4.)*(Teff/5772.)^(3./2.) 
   mcalc_org=mcalc ;mass with solar reference

   dif=mcalc
   cntc=0
   while abs(dif) GT 0.001 do begin
    mcalc0=mcalc
    refcalc=A1+A2*mcalc+A3/numax+A4*sqrt(numax)-A5*numax-A6*feh
    mcalc=(numax/3050.)^3.*(dnu/refcalc)^(-4.)*(Teff/5772.)^(3./2.) ;mass using Eq.3
    dif=mcalc-mcalc0
    cntc=cntc+1
   endwhile

  ; iterative computation for mass with Eq.4
   pubref=(0.64*feh+1.78)*exp(teff*(-0.55*feh+1.23)/10000.)*(cos(22.21*teff/10000.+(0.48*feh+0.12)))+0.66*feh+134.92
   mresc=(numax/3050.)^3.*(dnu/pubref)^(-4.)*(Teff/5772.)^(3./2.) ;mass using Paper~I
   mres_org=mresc

   dif=mresc
   cntref=0
   while abs(dif) GT 0.001 do begin
    mresc0=mresc
    rescalc=B1*mresc+B2*numax+(B3*mresc-B4*feh)/numax-B5-B6*mresc*numax
    mresc=(numax/3050.0d)^3.0d*(dnu/(rescalc+pubref))^(-4.d)*(teff/5777.d)^(3.d/2.d) ;mass using Eq.4
    dif=mresc-mresc0
    cntref=cntref+1
   endwhile

   ; computation of the radii
   rcalc_org=(numax/3050.)*(dnu/135.1)^(-2)*(Teff/5772.)^0.5 ;radius using solar references
   rcalc=(numax/3050.)*(dnu/refcalc)^(-2)*(Teff/5772.)^0.5   ; radius using Eq.3
   rres_org=(numax/3050.)*(dnu/(pubref))^(-2)*(Teff/5772.)^0.5    ;radius using Paper~I
   rresc=(numax/3050.)*(dnu/(rescalc+pubref))^(-2)*(Teff/5772.)^0.5  ;radius using Eq.4

  END
\end{lstlisting}


\bsp	
\label{lastpage}
\end{document}